%% file: main.tex
\documentclass[a4paper,11pt]{article}
\pdfoutput=1
\usepackage{jheppub} 
\usepackage{lineno}

\usepackage{amsmath}
\usepackage{amsfonts}
\usepackage{amssymb}
\usepackage{mathrsfs}
\usepackage{color}
\usepackage[dvipsnames]{xcolor}
\usepackage{slashed}
\usepackage{graphicx}
\usepackage{longtable}

\usepackage{multirow}
\usepackage{upgreek}
\usepackage[usestackEOL]{stackengine}
\strutlongstacks{T}

\usepackage[capitalise]{cleveref}
\usepackage[colorinlistoftodos]{todonotes}
\usepackage{isotope}

\usepackage{CJKutf8}

\usepackage{siunitx}
\input{units}

\usepackage{tikz}
\usepackage{tikz-feynman}
\input{tikz_style}


\arxivnumber{2211.02610} 

\title{New Clues About Light Sterile Neutrinos:  Preference for Models with Damping Effects in Global Fits}

\author[a]{J.M.~Hardin}
\author[b]{I.~Martinez-Soler}
\author[a]{A.~Diaz}
\author[b]{M. Jin (靳淼辰)}
\author[a]{N.W.~Kamp}
\author[b]{C.A.~Arg{\"u}elles}
\author[a]{J.M.~Conrad}
\author[c]{M.H.~Shaevitz}
\affiliation[a]{Dept.~of Physics, Massachusetts Institute of Technology, Cambridge, MA 02139, USA}
\affiliation[b]{Dept.~of Physics, Harvard University, Cambridge, MA 02138, USA}
\affiliation[c]{Dept.~of Physics, Columbia University, New York, NY, 10027, USA}

\emailAdd{jmhardin@mit.edu}

\abstract{This article reports global fits of short-baseline neutrino data to oscillation models involving light sterile neutrinos.
In the commonly-used 3+1 plane wave model, there is a well-known 4.9$\sigma$ tension between data sets sensitive to appearance versus disappearance of neutrinos.
We find that models that damp the oscillation prediction for the reactor data sets, especially at low energy, substantially improve the fits and reduce the tension.
We consider two such scenarios.
The first scenario introduces the quantum mechanical wavepacket effect that accounts for the source size in reactor experiments into the 3+1 model.
We find that inclusion of the wavepacket effect greatly improves the overall fit compared to a three-neutrino model by $\Delta \chi^2/\textrm{dof}=61.1/4$ ($7.1\sigma$ improvement) with best-fit $\Delta m^2=1.4~\textrm{eV}^2$ and wavepacket length of 67 fm.
The internal tension is reduced to 3.4$\sigma$.   If reactor-data only is fit, then the wavepacket preferred length is 91 fm ($>20$ fm at 99\% CL).
The second model introduces oscillations involving sterile flavor and allows the decay of the heaviest, mostly sterile mass state, $\nu_4$.
This model introduces a damping term similar to the wavepacket effect, but across all experiments.
Compared to a three-neutrino fit, this has a $\Delta \chi^2/\textrm{dof}=60.6/4$  ($7\sigma$ improvement)  with preferred $\Delta m^2=1.4~\textrm{eV}^2$ and decay $\Gamma = 0.35~\textrm{eV}$. The internal tension is reduced to 3.7$\sigma$.   

For many years, the reactor event rates have been observed to have structure that deviates from prediction.
Community discussion has focused on an excess compared to prediction observed at 5 MeV; however, other deviations are apparent.
This structure has $L$ dependence that is well-fit by the damped models.
Before assuming this points to new physics, we urge closer examination of systematic effects that could lead to this $L$ dependence.}

\begin{document}
\begin{CJK*}{UTF8}{gbsn}
\maketitle
\flushbottom
\end{CJK*}

\section{Introduction}

Since 1995, a series of experiments searching for neutrino oscillations have reported potential oscillation signals with $2\sigma$ to $5\sigma$ significance with mass-squared splittings, $\Delta m^2$, of $\mathcal{O}(1$ to 10~eV$^2)$.
These results do not fit our present picture of neutrino oscillations, which involves only three neutrinos with two mass splittings that are presently measured to be \SI{7.4e-5}{\eV\squared} and \SI{2.5e-3}{\eV\squared}~\cite{Esteban_2020}.
Thus, taking the results at face value, these results represent anomalies that may be pointing to new physics.
The simplest new physics solution introduces one additional neutrino that does not interact with the $W$ and $Z$ bosons,  and hence is called ``sterile''~\cite{Workman:2022ynf} but does participate in oscillations.
This model is called 3+1.

For the last decade, the 3+1 model has been challenged in multiple ways.
First, a set of experiments following  up on the anomalies have excluded signals due to the 3+1 scenario.
Second, global fits to the full data set have shown internal inconsistencies, commonly called ``tension.''~\cite{Kopp:2013vaa,Gariazzo:2017fdh,Diaz:2019fwt}
As a result, most of the community has reached the view that the simple 3+1 model cannot explain the data~\cite{Acero:2022wqg}.

This article presents global fits that look beyond simple 3+1 models.
In particular, we explore expanding beyond the plane-wave description of simple 3+1 to include ``wavepacket effects'' that take into account the finite size of the source producing the neutrino, as discussed in Refs. ~\cite{Arguelles:2022bvt, Jones:2014sfa, Jones:2022cvh, Akhmedov:2022mal, BenJosh, Banks:2022gwq, deGouvea:2020hfl, deGouvea:2021uvg}, and the references therein.
These effects show up as decoherence in experiments that have a relatively long $L/E$.
In the case of the experiments in the global fits, this will most affect reactors, damping the oscillation prediction at low energy.
Here, we show that this model, called ``3+1+WP'', to distinguish from the plane-wave 3+1 model, considerably reduces the tension in global fits.
We show that this solution improves the fits more than models that introduce additional sterile neutrinos, 3+2 or 3+3.

We also revisit a more complex model that has been examined in the past that introduces decay of the highest mass state in a 3+1 model, $\nu_4$~\cite{Palomares-Ruiz:2005zbh,Moss:2017pur,Diaz:2019fwt}.
The concept behind ``3+1+dk'' models is that while the sterile neutrino does not carry Standard Model couplings, it may carry Beyond Standard Model couplings to other new particles and/or to Standard Model particles.
In this case, one can expect decay, which will most affect the highest mass state due both to the larger mass and to the large sterile content from the mixing matrix.
Decay of the $\nu_4$ state damps oscillations at the largest mass splitting, and this affects all data sets, not just the reactor data sets.
However, we show that the best fit 3+1+dk parameters damp the reactor results while not substantially changing the prediction for other data sets leading to a very similar result as 3+1+WP. 

The work presented here builds on 15 years of global analyses~\cite{Collin:2016aqd,Conrad:2012qt,Sorel:2003hf,Moulai:2019gpi}, and especially our analysis from 2019~\cite{Diaz:2019fwt}.
The 2019 paper provides a pedantic review of the theory and practice of sterile neutrino fits; therefore, in Sec.~\ref{sec:models}, we describe the limitations of and extensions to the simple 3+1 model only briefly.
In Sec.~\ref{sec:Inputs2fits}, we focus on the latest information that has been included in the results, which involves several new data sets and some substantial updates.
In Sec.~\ref{sec:quality}, we only briefly discuss the methods of fitting and expressing the quality of fits, and the reader should see Ref.~\cite{Diaz:2019fwt} for details. 
Throughout this work, we are agnostic to the origin of the sterile neutrino masses and how they fit in an extended Standard Model.
On this latter point see Ref.~\cite{Montero:2022prj}. 
This has important implications in the experiments and bounds we consider in our work, since neutrinos in stellar environments or in the early Universe could have different properties.
For example, Ref.~\cite{Davoudiasl:2023uiq} recently proposed that the sterile neutrino mass originates from the interaction with an ambient field.  
This model makes the constraints from solar neutrinos not applicable to the scenario we discuss here--the terrestrial results are unchanged.   This is a similar to the questions associated with the cosmological data and represents an assumption of this analysis.

Additionally, cosmological data severally limits the 3+1-like scenarios by constraining the number of relativistic degrees of freedom and total sum of the neutrino masses; see~\cite{Acero:2022wqg} for a recent review.
One way to address the discrepancy between cosmological observables, like CMB and LSS, and eV-scale neutrinos is by considering alternative cosmological scenarios~\cite{Gelmini:2004ah,Hamann:2011ge}. Another solution is to introduce new forces that affect neutrinos, which are referred to as ``secret forces'' and have been discussed in recent studies~\cite{Song:2018zyl,Chu:2018gxk}.
These forces limit the production of sterile neutrinos in the early Universe before neutrino decoupling but result in mixing angles of around $\mathcal{O}(0.1)$ at present times.
Additionally, as recently shown in Ref.~\cite{Esteban:2022rjk} if long range neutrino forces exist cosmological constraints on the neutrino mass can be completely avoided.

Although the reactor experiment data have only been modestly updated since 2019, in Sec.~\ref{sec:Inputs2fits} we extend a discussion begun in Ref.~\cite{Diaz:2019fwt} concerning whether features in the reactor spectrum can obscure fit results; see Ref.~\cite{Coyle:2022bwa} for a similar recent discussion in the context of accelerator neutrinos.
Attention has focused on one feature of the reactor event energy spectrum, which is an excess at 5~MeV; however, the entire spectrum suffers many more excesses and deficits that vary from experiment to experiment.
The reactor community has taken ratios in various ways to reduce systematic pulls in sterile neutrino fits from these effects. In spite of that, we will show that suspicious structures that may not be related to Beyond Standard Model effects remain and are not identical between data sets and are not well covered by the systematic uncertainty provided in data releases for global fits.
We flag this because this may explain why damped models like 3+1+WP and 3+1+dk may be preferred.

In Sec.~\ref{sec:results}, we present the latest global-fit results.
These will show a preference for damped models for reactors.
We show that these models substantially improve the well-known tension between the appearance data subset and disappearance data subset.
Though some tension remains and we will show that the bulk of this tension can be attributed to the MiniBooNE data set.
As we note in Sec.~\ref{sec:Inputs2fits}, the MiniBooNE data is well-known for appearing to have two contributions to the excess, one of which matches the kinematic expectations for $\nu_e$ scattering, as expected from oscillations, and one of which is forward peaked~\cite{Aguilar-Arevalo:2020nvw}.

Finally, in Sec.~\ref{sec:discuss}, we revisit the information summarized above with more detail, in light of the quantitative results of Sec.~\ref{sec:results}.   

\section{Sterile Neutrino Models  \label{sec:models}}

The neutrino-extended Standard Model has three active neutrino flavors engaging in oscillations.
This is referred to as the ``null'' model in this study.
We will compare the models discussed below involving sterile neutrinos to the null model.
In this section, we introduce the examples used in our fits.
We also discuss metrics for comparing models. 

\subsection{Frequently Used Sterile Neutrino Models:  Plane-wave ``3+1'' and ``3+N''}

The community standard for comparing sterile neutrino searches has used the 3+1 model.
In this model, the Pontecorvo-Maki-Nakagawa-Sakata (PMNS) mixing matrix is expanded by one row and one column to accommodate the new flavor state (``s'') and mass state (``4''):
\begin{equation}
U_{3+1} = \begin{bmatrix}
U_{e1} & U_{e2} & U_{e3} & U_{e4} \\ 
\vdots & & \vdots & U_{\mu4} \\
\vdots & & \vdots & U_{\tau4} \\
U_{s1} & U_{s2} & U_{s3} & U_{s4}
\end{bmatrix}. \label{4mixmx}
\end{equation}
The ``short-baseline approximation'' is also frequently used, where $\Delta m_{41}^2 \gg |\Delta m_{31}^2| > \Delta m_{21}^2 $ is assumed and that the measured $L/E$ is small enough that only the oscillation due to $\Delta m_{41}^2$ is observable. 
The result is that the oscillation probabilities for $\nu_e$ disappearance, $\nu_\mu$ disappearance, and $\nu_e$ appearance are connected:
\begin{eqnarray}
P_{\nu_e \rightarrow \nu_e}&=&1-4(1-|U_{e4}|^2)|U_{e4}|^2\sin^2(\Delta_{41}L/E),~ \label{PUe4}\\
P_{\nu_\mu \rightarrow \nu_\mu}&=&1-4(1-|U_{\mu 4}|^2)|U_{\mu 4}|^2\sin^2(\Delta_{41}L/E),~ \label{PUmu4}\\
P_{\nu_{\mu}\rightarrow\nu_e}&=& 4|U_{e4}|^2 |U_{\mu4}|^2\sin^2(\Delta_{41}L/E),\label{PUe4mu4}
\end{eqnarray}
where $\Delta_{41}=1.27\Delta m^2_{41}$, in eV$^2$.  In this case, $L$ and $E$ are in units of m and MeV (or km and GeV), respectively. 
A consistent 3+1 global fit requires signals in all three of these oscillation modes with the same large mass-square splitting, $\Delta m^2_{41}$.
One also sees that the probabilities of appearance and disappearance must be related through a consistent set of matrix elements.  

The above equations are often re-written by replacing the combinations of matrix elements with ``pseudo''-mixing-angles:
\begin{eqnarray}
P_{\nu_e \rightarrow \nu_e}&=& 1-\sin^2 2\theta_{ee}\sin^2(\Delta_{41}L/E)~, 
\label{Pthetaee} \\
P_{\nu_\mu \rightarrow \nu_\mu}&=& 1-\sin^2 2\theta_{\mu\mu}\sin^2(\Delta_{41}L/E)~, \label{Pthetamumu} \\
P_{\nu_{\mu}\rightarrow\nu_e}&=& \sin^2 2\theta_{\mu e}\sin^2(\Delta_{41}L/E). \label{Pthetaemu}
\end{eqnarray}
In this article, we will present the results of the global fits within the context of these pseudo-mixing-angles.

Although 3+1 is the standard that is used for searching for new physics in short-baseline experiments, it is chosen for its ease of application to fits rather than because there is strong motivation to have only one sterile neutrino.
A less minimal model expands the sterile neutrino model to include a second sterile neutrino in the model (``3+2'').
Immediately, $4$ new parameters are added to the theory.
Fits to the most ``natural'' case, with three sterile neutrinos (``3+3'') become even more complicated, with a further $5$ parameters added to the theory.
The equations for 3+2 are provided in Refs.~\cite{Diaz:2019fwt},~\cite{Conrad:2012qt}.  The 2019 results show that 3+2 does not substantially improve matters. 

\subsection{Including Wavepacket Effects (``3+1+WP'')}
\label{sec:WP}

\begin{figure}[tb!]
 \includegraphics[width=0.7\textwidth]{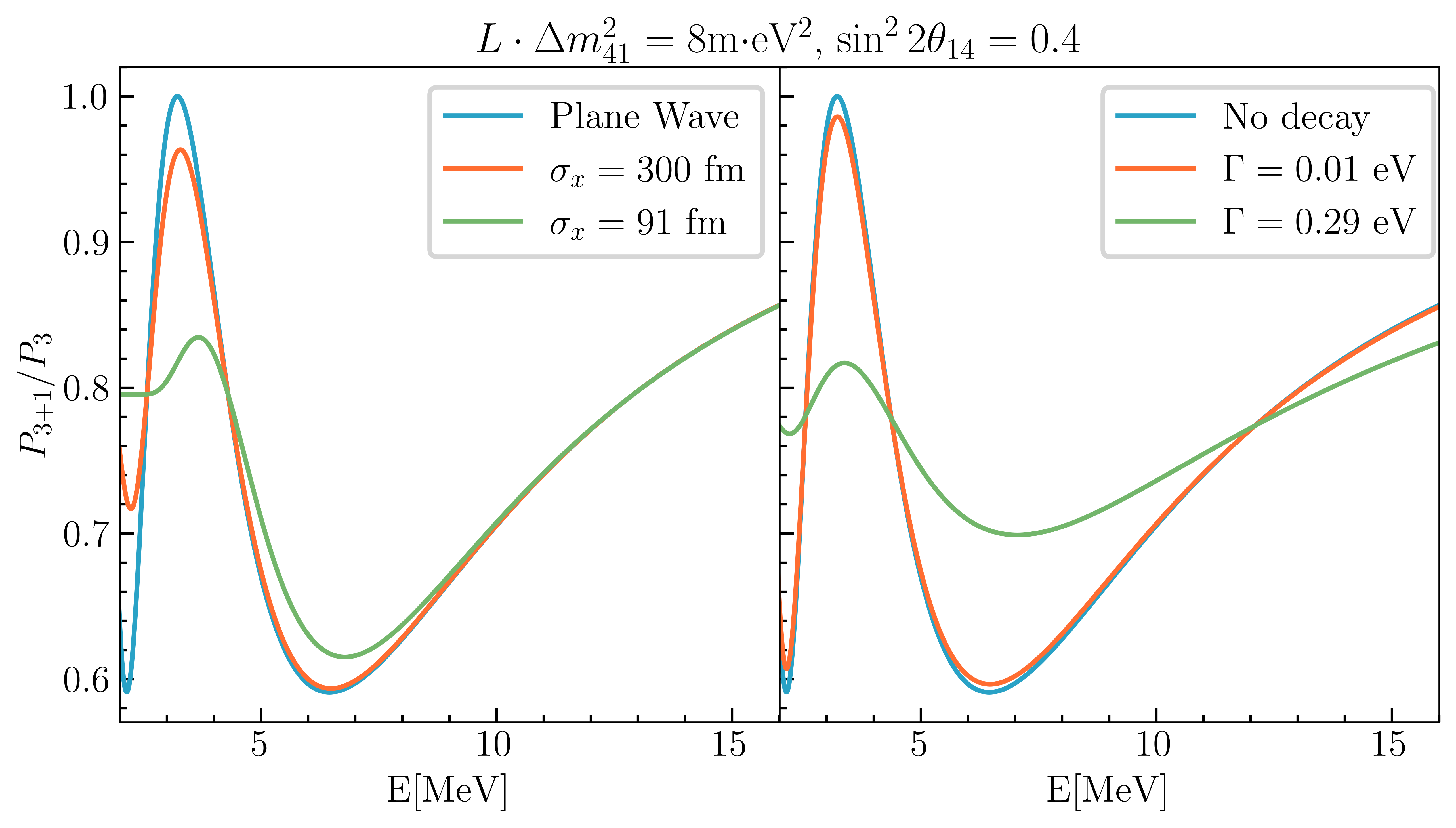}
\centering
\caption{Illustration of the damping of oscillations from the wavepacket effect (left) and from decay (right). The ratio of 3+1 to three neutrino oscillation probabilities is shown as a function of energy for fixed $L \Delta m^2$, which for a reactor experiment at $L=\SI{8}{\m}$ corresponds to $\Delta m^2=\SI{1}{\eV\squared}$--a case relevant to this study.  3+1, which is a plane-wave model, is shown in blue.  The orange and green curves are examples of modest and extreme damping at low energy for the 3+1+WP and 3+1+dk models.\label{fig:damplo}}
\end{figure}

\begin{figure}[tb!]
 \includegraphics[width=0.7\textwidth]{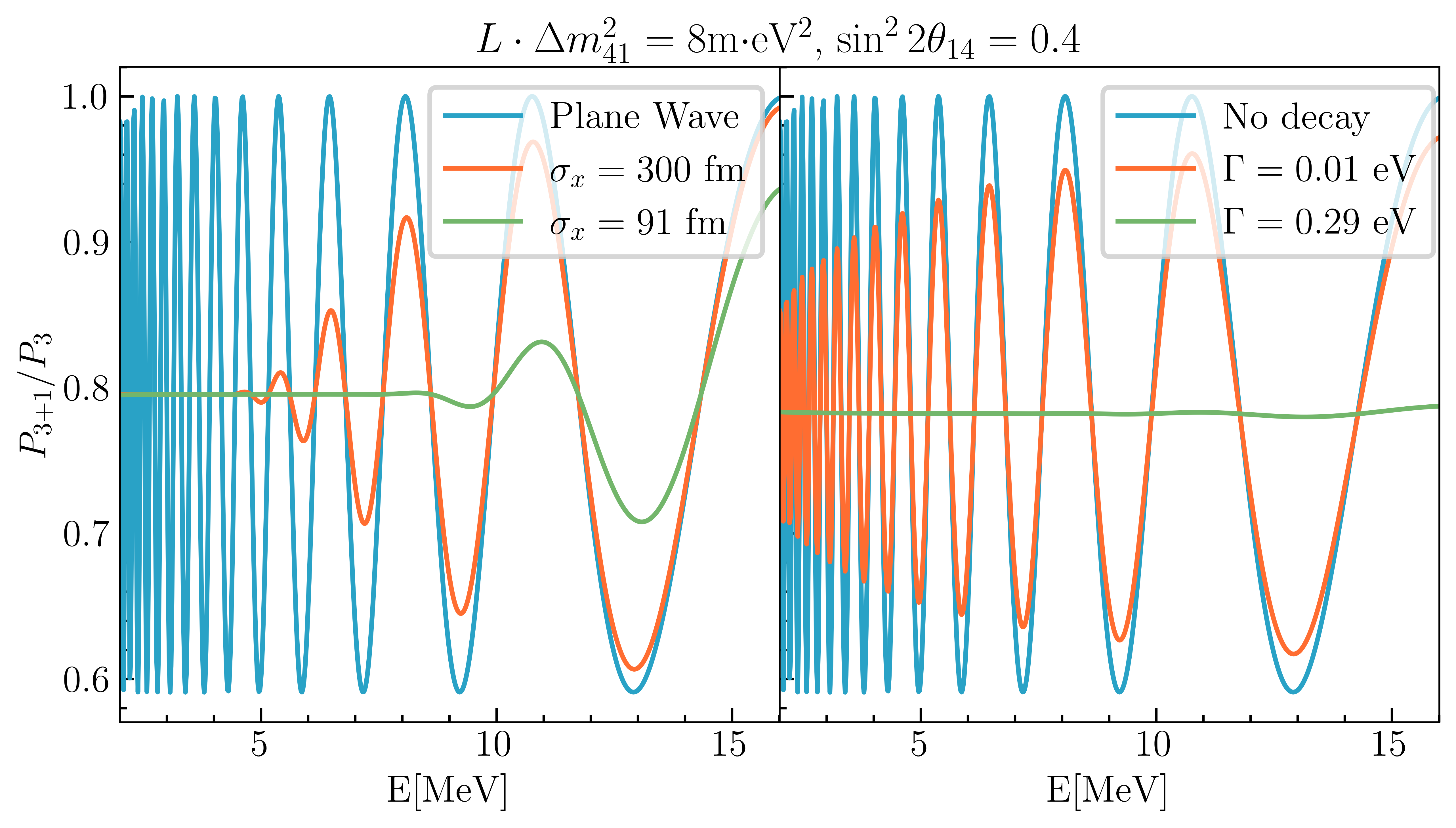}
\centering
\caption{Illustration of the damping of oscillations from the wavepacket effect (left) and from decay (right), as with Fig.~\ref{fig:damplo}, but for $\Delta m^2=10$ eV$^2$.  In this case, damping occurs across all relevant energies. \label{fig:damphi}}
\end{figure}

The 3+N models describe neutrinos as plane waves, where the interference between the massive states happens at any point in the space along the neutrino trajectory.
That description of the neutrino evolution is simplistic because it does not take into account that neutrinos are localized particles produced by a source of finite size.
This potentially introduces a ``wavepacket effect'':  as the neutrino propagates, the difference in the group's velocity of each massive state will result in a separation between them.
After some distance, the mixing between the massive states will stop, the oscillation will cease and the evolution can be described as an incoherent superposition of massive states.
The distance where the decoherence starts having an effect on the oscillation is given by the decoherence length~\cite{Arguelles:2022bvt}:
\begin{equation}
    L^{\text{coh}}=0.566 {\rm m} \left(\frac{E}{\rm MeV}\right)^2 \left(\frac{\sigma_x}{\rm 100 fm}\right) \left(\frac{\rm eV^2}{\Delta m^2}\right),\label{eq:lcoh}
\end{equation}
that depends on the neutrino energy ($E$ in MeV), the characteristic size of the wave packet ($\sigma_x$ in 100~fm), and the squared neutrino mass difference ($\Delta m^2$ in eV$^2$).    
The effect manifests as a characteristic ``damping effect'' at low values of the energy distribution if $L \Delta m^2$ is much larger than $E$.  

The wavepacket effect is a Standard Model nuclear effect that will show up within any oscillation data set at some level, though it may be too small to be observed.
Among today's three-neutrino oscillation experiments, the set that is expected to be most affected is reactor experiments, where the characteristic length of the U and Pu nuclei ($\mathcal{O}$(10~fm)) and the interatomic spacing ($10^{5}$~fm) each may contribute.

In a 3+1 model, this effect is included following the method of Ref.~\cite{Arguelles:2022bvt}, with the electron-flavor disappearance probability modified to be:
\begin{eqnarray}
    P_{\nu_e \rightarrow \nu_e}^{WP} &=& 1 - \sin^2 2\theta_{ee} \Big[ \big( {{1-e^{-A^2}}\over{2}} \big) \nonumber \\
    &&+\sin^2 (\Delta_{41} L/E) e^{-A^2} \Big]~,
\end{eqnarray}
where $A= L/L_{\text{coh}}$ and $L_{\text{coh}}$ are proportional to $\sigma_x$ as seen in eq.~\ref{eq:lcoh}.
If $\sigma_x$ is sufficiently small, the effect becomes observable.  

Figs.~\ref{fig:damplo} and \ref{fig:damphi}, left, illustrate the effect of varying $\sigma_x$ on the predicted signal for a reactor experiment located at $L=\SI{8}{\m}$, for $\Delta m^2=\SI{1}{\eV\squared}$ and $\SI{10}{\eV\squared}$, respectively.
The predictions for the models are shown as a ratio to the null model.
The blue line is a 3+1 (plane wave) model.
The orange and green lines show examples of the effect for two different $\sigma_x$.
Fig.~\ref{fig:damplo} shows that at $\Delta m^2=1$~eV$^2$, the damping occurs at low energy, while Fig.~\ref{fig:damphi} shows that at $\Delta m^2=10$~eV$^2$, the damping occurs across the entire relevant energy range for reactors.

Following Ref.~\cite{Arguelles:2022bvt}, wavepacket effects can potentially be significant for large nucleus source experiments such as BEST, GALLEX, and SAGE.  We choose to fit those datasets for the same $\sigma_x$ as the reactors.  However, in agreement with Ref.~\cite{Arguelles:2022bvt}, the wavepacket effect on these results turns out to be minimal due to the experimental design.  These are ``counting experiments'' that effectively integrate over $L/E$, leading to little discrimination between damped and 3+1 models.
We note that for production from pion/muon decay beams, decoherence effects are expected to be negligible due to a combination of the small size of the meson, point-like nature of the muon, and the high energy of these experiments \cite{Jones:2014sfa}.

We do not choose to consider any theoretical prediction for $\sigma_x$~\cite{Akhmedov:2022bjs, Jones:2022cvh, Akhmedov:2022mal, BenJosh}, and instead investigate this question in an agnostic, experiment-driven manner as in Ref.~\cite{Arguelles:2022bvt}.  We encourage the reader to see Ref.~\cite{Banks:2022gwq} and references therein for beyond Standard Model mechanisms that effectively change the neutrino wavepacket size.

Specifically, we will perform global fits to the 3+1+WP, applying the wavepacket scenario as described above, allowing $\sigma_x$ to float.
We will then compare that result to ``only 3'' fits from Refs.~\cite{deGouvea:2020hfl, deGouvea:2021uvg}, with the caveat that this comparison is not apples to apples due to the additional neutrino state in our fits.
As discussed below, our findings are that a small value of $\sigma_x$, which causes strong damping for the reactor results (green case in Fig.~\ref{fig:damphi}), is preferred.

\subsection{3+1 with decay (``3+1+dk'')}

The wavepacket effect is not the only way to introduce damping into the 3+1 picture.
A second model that causes damping allows neutrino decay, which we call ``3+1+dk''.
The phenomenological motivation for 3+1+dk is that the existence of neutrino mass also permits decay of the mass state if the ``sterile'' neutrino has Beyond Standard Model interactions.
The possibility of decay of $\nu_4$, the primarily-sterile state, has been considered in many references~\cite{Ma:2001ip,PalomaresRuiz:2005vf,Gninenko:2009ks,Dib:2011jh,Masip:2012ke,Masip:2011qb,Gninenko:2012rw,Moss:2017pur,Ballett:2018ynz,Ballett:2016opr,Fischer:2019fbw,Dentler:2019dhz,deGouvea:2019qre,deGouvea:2021ual}.
Sterile neutrino decay leads to a damping in the neutrino oscillation patterns in $L/E$, as well as production of new Beyond Standard Model particles and of known particles such as photons, or of $\nu_1$, $\nu_2$ and $\nu_3$, thereby regenerating the flux.
In this article, we will only consider the damping effect on oscillation signals with decay to Beyond Standard Model particles.  In this case, called ``invisible decay,'' the $\nu_4$ component of the flux dies off with distance travelled.  

The experimental motivation for 3+1+dk came from analysis of IceCube $\nu_\mu$ disappearance data~\cite{PhysRevLett.129.151801}.
The IceCube collaboration found that their atmospheric $\nu_\mu$ data were fit better with 3+1+dk than 3+1, and that 3+1 was a slightly better fit than null.
An important point of departure from the 3+1+WP model is that 3+1+dk is applied to all data sets, not just the reactor data set.
The survival probability for the $\nu_e$ disappearance case is given by:
\begin{eqnarray}
P^{\rm 3+1+dk}_{{\nu}_e \rightarrow {\nu}_e}&&= \nonumber \\
&&    2 |U_{e4}|^{2} e^{-2.53 \frac{m_4  L}{\tau E}} (1 - |U_{e4}|^{2})\cos\left(2.53 \Delta m^2_{41} (L/E\right) \nonumber \\
&& + |U_{e4}|^{4} e^{-5.07 \frac{m_4 L}{\tau E}} + (1 - |U_{e4}|^{2})^{2}~, \label{3+1+dk}
\end{eqnarray}
where $m_4=\sqrt{\Delta m^2_{41}}$, since the smallest mass state is assumed to have negligible mass.
The other oscillation probabilities are modified in a similar manner.
The decay causes the $\nu_4$ state to die away as a function of $L/E$, so experiments with relatively large $L/E$, like the reactor experiments, become insensitive to oscillations.   

We illustrate the 3+1+dk effect on reactors in Figs.~\ref{fig:damplo} and \ref{fig:damphi}, right.
As described for the wavepacket illustration, the orange and green represent modest ($\Gamma=\SI{0.01}{\eV}$) and extreme ($\Gamma=\SI{0.29}{\eV}$) values of the damping parameter.

\section{Inputs to the Global Fits \label{sec:Inputs2fits}}

Table~\ref{table:data} maps all data sets used in this analysis to the type of fit that applies.   
Compared to the fits presented in Ref.~\cite{Diaz:2019fwt}, the fits in this article include new experiments and updates to previous experiments.
The experiments added were: BEST~\cite{Barinov:2022wfh},  STEREO~\cite{PhysRevD.102.052002}, and MicroBooNE \cite{MicroBooNE:2021nxr}.
The MiniBooNE \cite{MiniBooNE:2022emn}, MINOS+~\cite{MINOS:2008hdf}, and IceCube~\cite{Aartsen:2020iky} results have been substantially updated.
The remaining citations for experiments that are not updated, and are described in detail in Ref.~\cite{Diaz:2019fwt}, are Refs.~\cite{MBnubar,MBNumi,NOMAD1,Aguilar:2001ty,KARMEN,SBMBnu,SBMBnubar,SAGE,Gallex, Bugey,DANSS,PhysRevD.103.032001}.
\footnote{We have not included data from Neutrino-4~\cite{Serebrov:2020kmd} because we were not able to reproduce their results based on the information that they have provided, and the authors did not respond to our questions.
The problems with treatment of Neutrino-4 data that we identified agrees with those observed in other studies~\cite{Giunti:2021iti}.}

\begin{table}[t]
\begin{center}
\begin{tabular}{| c || c | c | c |}
 \hline
  & $\nu_\mu \rightarrow \nu_e$ & $\nu_\mu \rightarrow \nu_{\mu}$ & $\nu_e \rightarrow \nu_e$ \\
 \hline\hline
 $\nu$ & \Centerstack{\centering MiniBooNE \\ NUMI-MB \\ NOMAD \\ MicroBooNE}&  \Centerstack{\centering SciBooNE-MB \\ CCFR \\ CDHS \\ MINOS \\ MicroBooNE} &  \Centerstack{\centering KARMEN-LSND-xsec \\ SAGE+GALLEX \\ BEST \\ MicroBooNE \\ MiniBooNE} \\  
 \hline
 $\bar{\nu}$ & \Centerstack{\centering LSND \\ KARMEN \\ MiniBooNE} & \Centerstack{\centering SciBooNE-MB \\ CCFR \\ MINOS \\ IceCube} & \Centerstack{\centering Bugey \\ NEOS \\ DANSS \\ PROSPECT \\ STEREO \\ MiniBooNE}  \\ 
 \hline
\end{tabular}
\caption{\label{table:data} The data sets used in global fits presented in this article, divided according to the type of fit applied.
Citations for data sets are provided in the text.}
\end{center}
\end{table}

\subsection{New to These Fits:  MicroBooNE}
\begin{figure}[tb!]
 \includegraphics[width=0.95\textwidth]{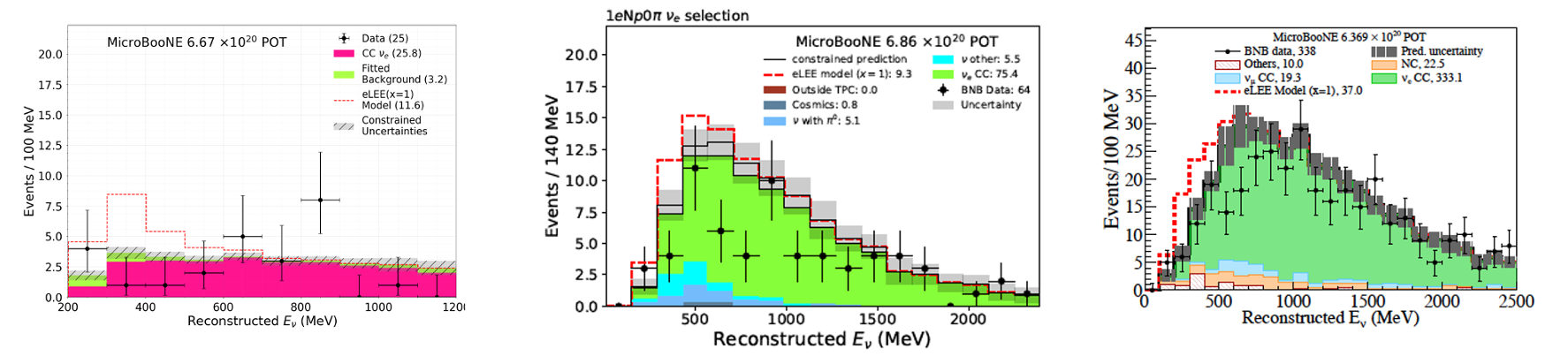}
\centering
\caption{MicroBooNE $\nu_e$ energy spectra from Ref.~\cite{MicroBooNE:2021nxr} from the CCQE analysis, left; semi-inclusive analysis, middle; and inclusive analysis, right. \label{fig:microbooneenergy}}
\end{figure}

The MicroBooNE experiment released first results probing the electron neutrino interpretation of the MiniBooNE excess in October 2021~\cite{MicroBooNE:2021nxr}.    
Three analyses looked for an excess of electron neutrinos in the Booster Neutrino Beam, focusing on (1) an exclusive sample of $\nu_e$ CCQE-like events with one electron and one proton in the final state, (2) a semi-inclusive sample of MiniBooNE-like $\nu_e$ CC events with no pions in the final state, and (3) an inclusive sample of all $\nu_e$ CC interactions.
The results were very surprising, indicating a deficit of $\nu_e$ with respect to the Standard Model~\cite{Denton:2021czb}, not an excess, as seen in in Fig.~\ref{fig:microbooneenergy}~\cite{MicroBooNE:2021nxr}. The $\nu_\mu$ data, on the other hand, is higher than prediction~\cite{MicroBooNE:2021nxr}.    

Subsequent fits by MicroBooNE ~\cite{MicroBooNE:2022sdp}, as well as others ~\cite{Arguelles:2021meu,Denton:2021czb, MiniBooNE:2022emn}, interpret the inclusive sample (Fig.~\ref{fig:microbooneenergy}, right) within a 3+1 model.   The combination of MicroBooNE's $\nu_e$ deficit and low statistics leads this data set to not significantly shift the MiniBooNE result in a combined two-experiment fit ~\cite{Arguelles:2021meu,MiniBooNE:2022emn}.  The $\sim 2\sigma$ deficit also weakened the $\nu_e$ disappearance limit from MicroBooNE compared to the sensitivity \cite{MicroBooNE:2022sdp}. 

A limitation in including MicroBooNE data in fits is that the publication from the 3+1 analysis \cite{MicroBooNE:2022sdp} did not include a data release.  Thus all external fits to the data can make use of only the released data from Ref.~\cite{MicroBooNE:2021nxr}.   In this study, we follow the procedure described in Ref.~\cite{MiniBooNE:2022emn} to address the missing information.
Specifically, we leverage the MiniBooNE Booster Neutrino Beam simulation to derive the prediction in four of the seven samples of the MicroBooNE inclusive analysis: $\nu_e$ and $\nu_\mu$ fully-contained and partially-contained events.   The limitation of this approach is that it does not include information from the remaining three $\pi^0$-based samples~\cite{hepdata.114862}.  Nevertheless, as originally seen in Ref.~\cite{MiniBooNE:2022emn}, this approach well-reproduces the published MicroBooNE limits \cite{MicroBooNE:2022sdp}.

For the statistical treatment of the inclusive analysis, we use the combined Neyman-Pearson $\chi^2_{\rm CNP}$~\cite{Ji:2019yca}.
This test statistic attempts to incorporate finite Monte Carlo statistics in the small sample size regime.  See Refs.~\cite{Barlow:1993dm,Chirkin:2013lya,Glusenkamp:2017rlp,Arguelles:2019izp,Glusenkamp:2019uir} for further discussion of this topic.
We use the joint covariance matrix provided by MicroBooNE to account for correlations between the four different samples~\cite{hepdata.114862}.

\begin{figure}
    \centering
    \includegraphics[width=0.45\textwidth]{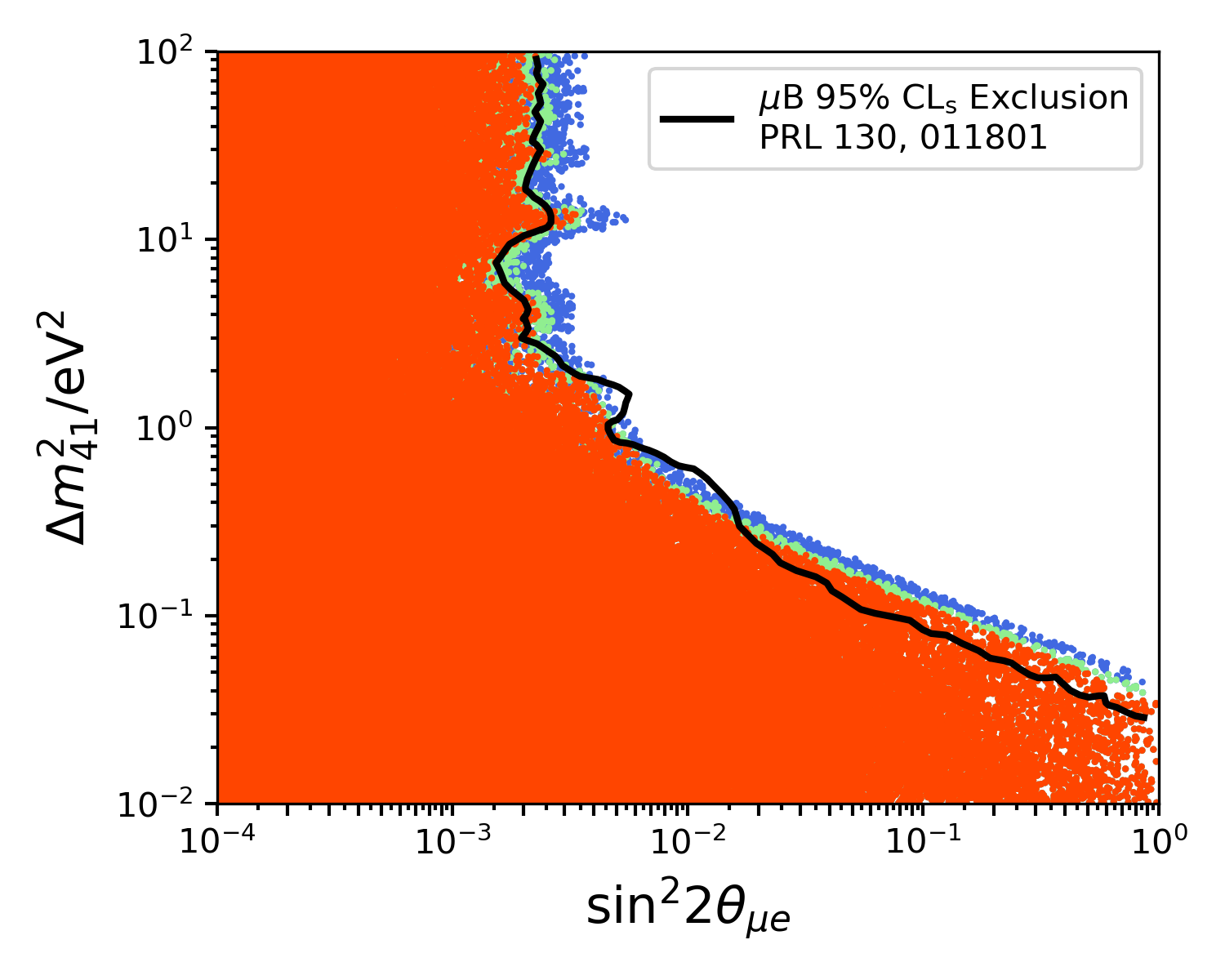}
    \caption{Constraints in $\Delta m_{41}^2$-$\sin^2 2\theta_{\mu e}$ parameter space calculated in this analysis (colored points) compared to constraints published by the MicroBooNE collaboration~\cite{MicroBooNE:2022sdp} (black line). The MicroBooNE constraints should be compared to the rightmost edge of the green points, which corresponds to our 95\% CL constraint in this parameter space.}
    \label{fig:wc_mue}
\end{figure}

We also consider two additional nuisance parameters to account for potential additional uncertainty in the relative efficiency of the $\nu_e$ and $\nu_\mu$ channels of the inclusive analysis~\cite{MiniBooNE:2022emn}.
This is motivated by systematic disagreement between data and prediction in the $\nu_\mu$ channels.
The introduced nuisance parameters adjust the overall normalization of the prediction in the $\nu_\mu$ FC and PC channels.
We consider a flat prior bounded within $\pm 1\sigma$ of the nominal prediction in the channel, where $\sigma$ is defined as the uncertainty on the overall normalization in each channel calculated from the reported covariance matrices.
The nuisance parameters are not penalized within the $\pm 1\sigma$ window and thus effectively re-scale the overall normalization in each $\nu_\mu$ channel under a given $3+1$ model hypothesis toward the observed data (bounded by $\pm 1\sigma$ of the nominal normalization).
This is important for the $\nu_\mu$ disappearance result, as the inclusive analysis sets much stronger constraints on $\sin^2 2\theta_{\mu \mu}$ compared to their sensitivity when these nuisance parameters are not included.
Introducing the parameters relaxes the observed constraint to more closely match the $\nu_\mu$ disappearance sensitivity calculated in Ref.~\cite{Arguelles:2021meu}.
We have also checked that our constraints in $\Delta m_{41}^2$-$\sin^2 2\theta_{\mu e}$ parameter space match the published MicroBooNE constraints~\cite{MicroBooNE:2022sdp}, as shown in Fig.~\ref{fig:wc_mue}.
It is important to note that the MicroBooNE collaboration uses the CL$_s$ method to account for fluctuations in the observed data with respect to the no-oscillation hypothesis~\cite{MicroBooNE:2022sdp}, a strategy which is not possible in our global fit.
One could in principle achieve the same effect through a proper Feldman-Cousins interpretation of the inclusive analysis statistical results; however, this is prohibitively difficult in the context of the current global fit.

\subsection{New to These Fits: BEST}

The Baksan Experiment on Sterile Transitions (BEST) experiment ran to follow up on previously observed anomalies in Gallium data~\cite{Barinov:2021asz}.
In 2019, a $(3.414\pm0.008)$ MCi \isotope[51]{Cr} source was placed in the center of a dual volume gallium detector.
The inner spherical volume of diameter 133.5~cm held 7.5~tons of Ga, while the outer cylindrical volume with dimension $(h,r)=(234.5, 109) $ cm held 40.0~tons. 
    
Like the previous Gallium anomalies, BEST observed a deficit in Ge production rate in both volumes, with ratios of data to prediction of $R_\textrm{in}=0.791\pm0.05$ and $R_\textrm{out}=0.766\pm0.05$.
The deficits are between $4\sigma$ and 5$\sigma$, which is very large compared to results from most electron-flavor experiments.
This new addition to our fits joins two other data sets with $>4\sigma$ signals, LSND and MiniBooNE.

While an overall deficit is observed in each volume, there is no clear oscillatory effect between volumes. 
However, for relatively high $\Delta m^2$ oscillations may be rapid compared to the $L$ in each volume, which would average out the signal.
Averaging due to rapid oscillations produces damping for large $L/E$.

In combination with the previous Gallium anomalies, a strong preference for the $\nu_e$ oscillation  signature is found with a large mixing angle of $\sin^2 2\theta = 0.34$ recovered for $\Delta m^2 \gtrsim 1\ \textrm{eV}^2$.

\subsection{Updated for These Fits: IceCube}

The sterile neutrino analysis with the IceCube Neutrino Observatory near the geographical South Pole studies neutrinos events that span an energy range from approximately 500~GeV to 10~TeV and examines the northern hemisphere.  
The IceCube analysis examines observed the $\nu_\mu$ event rate for upward, through-going interactions below the detector and contained interactions within the detector.
The data are analyzed as a function of reconstructed neutrino energy, which has a large smearing for events below the detector, and zenith angle, $\theta_z$.
The zenith angle plays the role of $L$ in the usual oscillation analysis.

A deficit in this data set would arise from matter-induced effects as well as vacuum oscillations, hence this data set is very different from the rest listed in Table~\ref{table:data}, which are all subject to only vacuum oscillations.
At TeV energies, in the presence of a light sterile neutrino, matter-enhanced resonances producing deficits of events are expected to appear in the energy vs. $\cos \theta_z$ event plane.
The location of the resonance disappearance depends on the mass-square difference, $\Delta m^2_{41}$ and mixing angle, $\theta_{24}$.
For small mixing angles the resonance happens for core-thoroughgoing trajectories, while for larger mixing they can extend to the mantle region.
The mass splitting controls the resonance location in energy, where the larger the difference, the higher the resonance energy.
See Ref.~\cite{Diaz:2022grj} for an extended discussion.

To date, the global fits where we have included IceCube results encompassed only 1 year of data~\cite{Dentler:2018sju,Moulai:2019gpi}.
However, recently IceCube has supplied data releases for eight years of data analyzed in a 3+1 scenario~\cite{IceCube:2020phf} and in a 3+1+dk scenario~\cite{IceCubeCollaboration:2022tso}.
These latest analyses used an improved event selection and systematic treatment described in Ref.~\cite{IceCube:2020tka}.
Notable improvements in systematic treatments include an improved treatment of the atmospheric flux uncertainties~\cite{Moulai:2021zey} and a new treatment of the bulk ice uncertainties~\cite{IceCube:2019lxi}.
To date, IceCube has only released likelihoods with these specific models and not bin-by-bin event rates compared to Standard Model.
As a result, this data release can only be interpreted within in 3+1, 3+1+WP (since for IceCube this will be degenerate with 3+1), and 3+1+dk.
The likelihoods are not available for 3+2 and 3+3 fits and are not included in our analysis.


\subsection{Updated for These Fits: MiniBooNE \label{sec:MB}}

\begin{figure}[tb!]
 \includegraphics[width=0.5\textwidth]{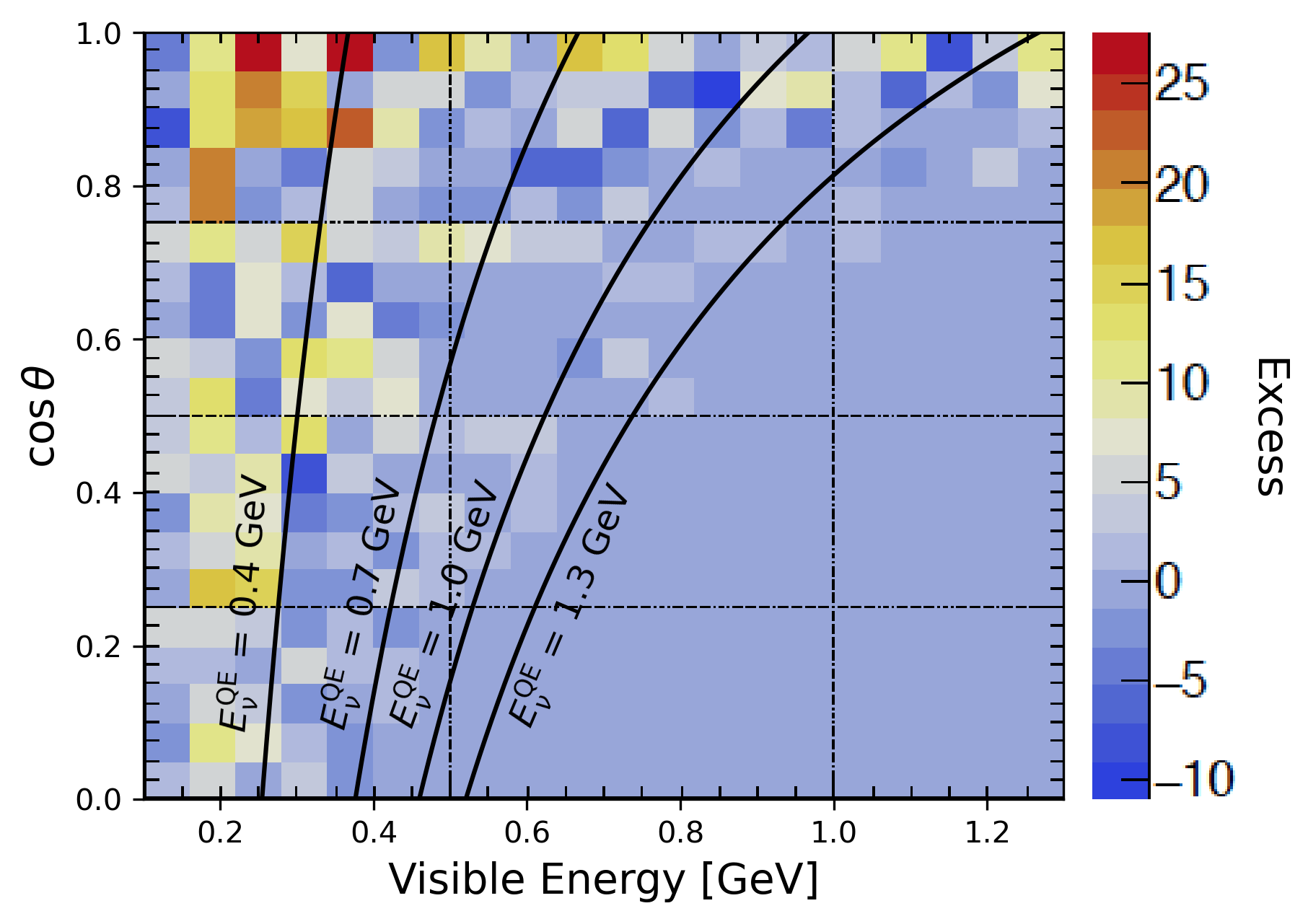}
\centering
\caption{The MiniBooNE excess is described as a function of $E_{vis}$ and $\cos(\theta)$, where $\theta$ is the electron scattering angle, from Ref. \cite{MiniBooNE:2020pnu}.  We have annotated the plot by adding lines of constant $E_\nu$.  If the sample were strictly due to $\nu_e$ charged current quasielastic scattering, the events would be distributed along these curves according to the neutrino energy prediction.  While true of about half the event sample, one notes a cluster of events at $\cos(\theta)>0.9$ and low energy that deviates from expectation.   The origin of this cluster is not yet identified.}
\label{fig:mbeventdistribution}
\end{figure}

For these global fits, we use the final MiniBooNE data set that was released in 2022~\cite{MiniBooNE:2020pnu}.
We have updated our analysis to account for all $3+1$-related phenomena, $\nu_e$/$\bar{\nu}_e$ appearance and disappearance as well as $\nu_\mu$/$\bar{\nu}_\mu$ disappearance, matching the treatment outlined in Ref.~\cite{MiniBooNE:2022emn}.
In Ref.~\cite{MiniBooNE:2020pnu}, the MiniBooNE collaboration noted that the $\nu_e$ excess in the data appears to have two contributions, one that follows the expectation of CCQE scattering and one that is at low energy and more forward peaked than expected.
The two contributions are indicated in Fig.~\ref{fig:mbeventdistribution}.

This observation has led to two explorations.  The first is the ``Altarelli Cocktail'' where a collection of systematic effects all combine~\cite{Brdar:2021ysi}.
The second is a mixed model involving oscillations and new physics which potentially explains an excess of photons at low-energy, where representative examples are Refs.~\cite{Vergani:2022qre,Kamp:2022bpt}.
For the fits here, we will assume an explanation for MiniBooNE that is fully due to oscillations.
However, we will show that MiniBooNE does contribute at a substantial level to the tension measured in the results, as discussed in Sec.~\ref{sec:WP}, and this would be consistent with either explanation.

\subsection{Updated for These Fits: MINOS+}

Long-baseline (LBL) experiments take advantage of measuring the neutrino flux from a well-defined direction and baseline to determine the oscillation parameters with good precision.
To reduce the uncertainties related to the flux and cross-section, LBL experiments typically use a near/far configuration, where the flux is measured at two different baselines.
In the case of MINOS+~\cite{MINOS:2008hdf}, the neutrino flux generated in the NuMI beam is measured in a near detector at a $1.04$~km baseline and in a far detector at $735$~km. 

The flux peaks at \SI{\sim 3}{\GeV}, but extends over a wide energy range from $0.5$~GeV to $40$~GeV, which allows constraints on $\Delta m^2_{31}$ and $\sin^2\theta_{23}$~\cite{MINOS:2014rjg} within the $3\nu$-mixing scenario.
In the $3+1$ scenario, the broad energy range makes it possible to study a large range of masses.
The oscillation length is comparable to the distance to the far detector for $\Delta m^2_{41}\in [10^{-3},10^{-1}]\text{eV}^2$.
The oscillation will happen at shorter distances for larger masses. For $\Delta m^2_{41}\in [1,50]\text{eV}^2$, the neutrino will show a flavor oscillation in the near detector. In the far detector, the new phase will be too large to be resolved by the detector resolution, and it will average out. For masses above $\sim 50$~eV$^2$, the oscillation length is shorter than the distance to the near detector, and the oscillations will also be averaged out in the near detector.  

The combined analysis between MINOS and MINOS+~\cite{MINOS:2017cae}, which correspond to an exposure of $16.36\times 10^{20}$ POTs, excludes $\sin^2\theta_{24} > 5\times 10^{-3}$ for masses between $ 10^{-2}~\text{eV}^2 < \Delta m^2_{41} < 10~\text{eV}^2$ at 90\% CL.  
Above $\Delta m^2 > 10~\text{eV}^2$ mixings larger than $\sin^2\theta_{24} > 2\times10^{-2}$ are excluded. 
Those results are the strongest bounds on sterile neutrinos using the muon-disappearance analysis. We have included them in our present analysis, although we note that questions have been raised about the unexpectedly strong limit of the MINOS/MINOS+ data set at high $\Delta m^2$ where the result should be normalization dominated \cite{Diaz:2019fwt}.

\subsection{Updated for these fits:  Reactors}

The inputs from reactor experiments have relatively modest updates compared to the 2019 fits~\cite{Diaz:2019fwt}.

The NEOS experiment is composed of a single detector.
In order to minimize systematic errors due to nuclear reactor models, they compare their data with a reference flux. 
At the time of the 2019 fits, NEOS used Daya Bay's unfolded $\bar{\nu}_e$ flux measurement~\cite{NEOS:2016wee}.
Since then, the NEOS collaboration conducted a joint analysis with the RENO collaboration~\cite{RENO:2020hva}.
Now, the unfolded $\bar{\nu}_e$ flux measurement from the RENO near detector is used as the reference flux.
The RENO near detector is placed at the same reactor complex as the NEOS detector, so this change reduces the systematic errors which may arise due to using different reactor cores.
The NEOS detector data, otherwise, remains unchanged.
We update the NEOS fits accordingly.

The PROSPECT data have been updated since the 2019 fit.
The previous data set had a reactor-on time of 33 days \cite{PROSPECT:2018dtt}, while the most recent data set has a reactor-time of 96 days \cite{PROSPECT:2020sxr}.

A more important update is that STEREO \cite{PhysRevD.102.052002} is now included in the global fits, which was not the case for the 2019 global fits because the data were not yet published.    However, the 2019 global fits updated with STEREO appeared in Ref.~\cite{Vergani:2022qre}.
The STEREO data are given in eleven energy bins across six different cells of increasing baseline, separated between two different run periods for a total of 179 days of reactor-on data.
The fit incorporates free normalization parameters for each energy bin, thus remaining agnostic to the reactor $\bar{\nu}_e$ flux while searching for oscillations as a function of baseline.
Following the recommendation of the STEREO collaboration~\cite{PhysRevD.102.052002}, we impose a 20\% uncertainty on the overall rate to compensate for the nonstandard $\Delta \chi^2$ distribution compared to the assumption from Wilks' theorem.

\subsection{Comments about the Reactor Data \label{sec:reactdiscuss}}

Overall, the reactor data sets deserve substantial discussion given their key role in these global fits.
These data sets exhibit structure that does not necessarily cancel in ratios and that may not be due to oscillations.
The structure varies from experiment to experiment.

\begin{figure}[tb!]
 \includegraphics[width=0.5\textwidth]{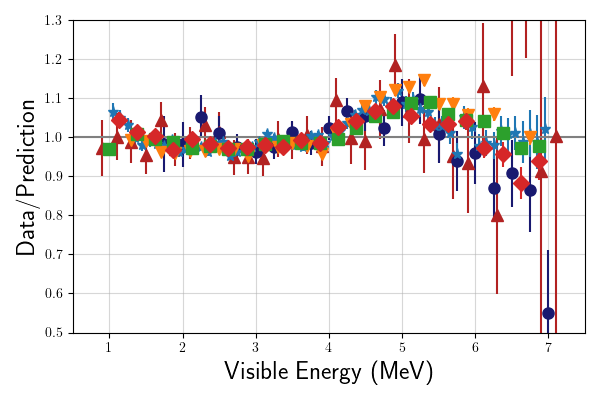}
\centering
\caption{Reactor data normalized by the predicted flux. Data are from RENO (inverted triangle) \cite{RENO}, Daya Bay (square) \cite{DayaBay:2016ssb}, Double Chooz (diamond) \cite {DC5MeV}, NEOS (stars) \cite{NEOS}, PROSPECT (triangle) \cite{PhysRevD.103.032001} and STEREO (solid dots) \cite{PhysRevD.102.052002}. \label{fig:mostreactors}}
\end{figure}

Let us begin by considering the measured reactor energy spectrum compared to the prediction.
This is summarized for many reactor experiments, not all of which are used in these global fits because of their baseline length, in Fig.~\ref{fig:mostreactors}.
One sees experiments running at power reactors (RENO, Daya Bay, Double Chooz, NEOS) and those running at research (HEU) reactors (PROSPECT, STEREO) have generally similar features.
A well-known 5 MeV excess (often called a ``bump'')  is observed.
What is not thoroughly discussed within the community is that other features also appear.
There is a deficit in some experiments above 6 MeV.
There is an excess in most experiments below 1.5 MeV.
The range between 1.5 and 4 MeV may have structure also.  While it is convenient to show many reactor data sets on one plot, additional individual spectral features are obscured.

\begin{figure}[tb!]
 \includegraphics[width=0.49\textwidth]{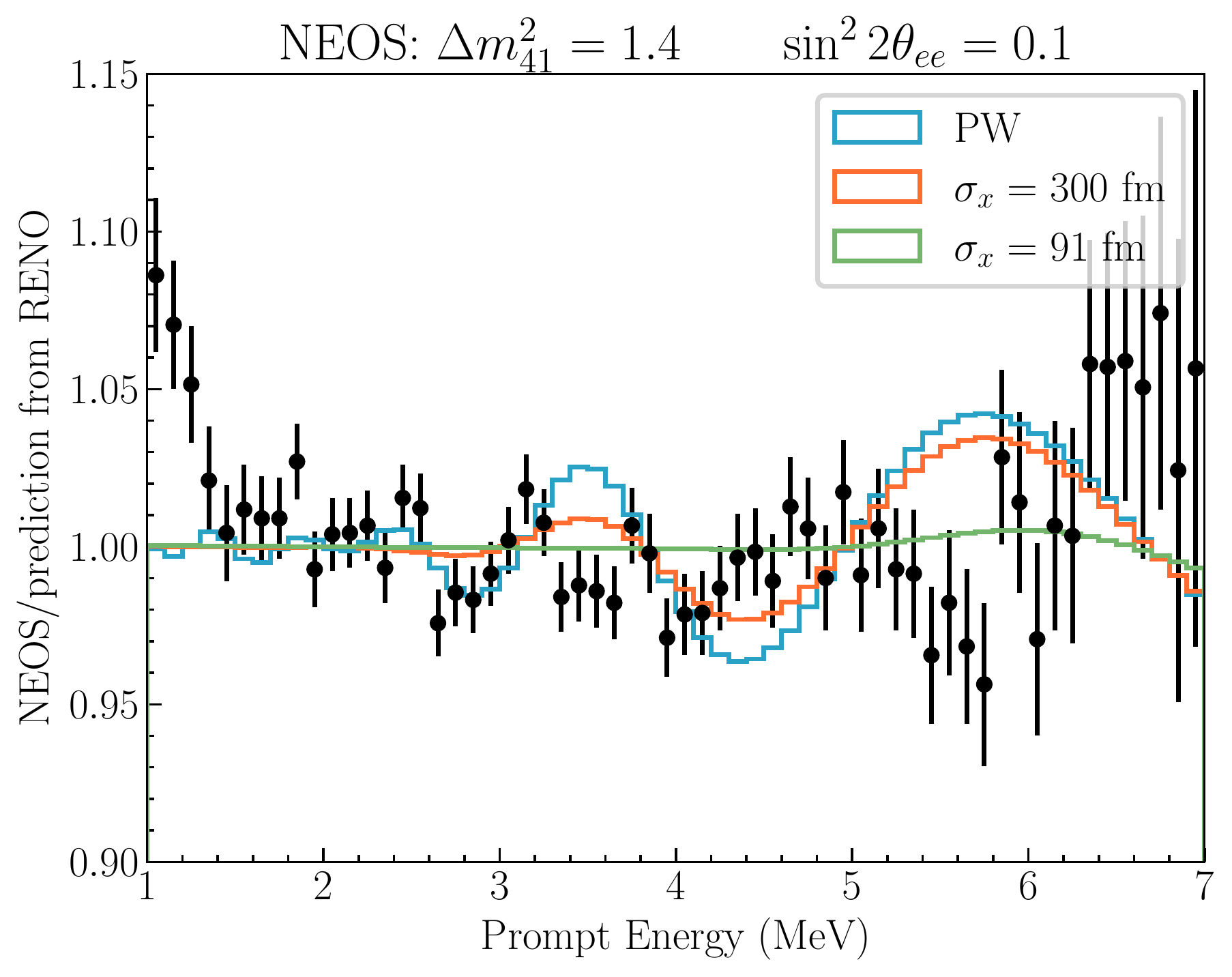}
  \includegraphics[width=0.49\textwidth]{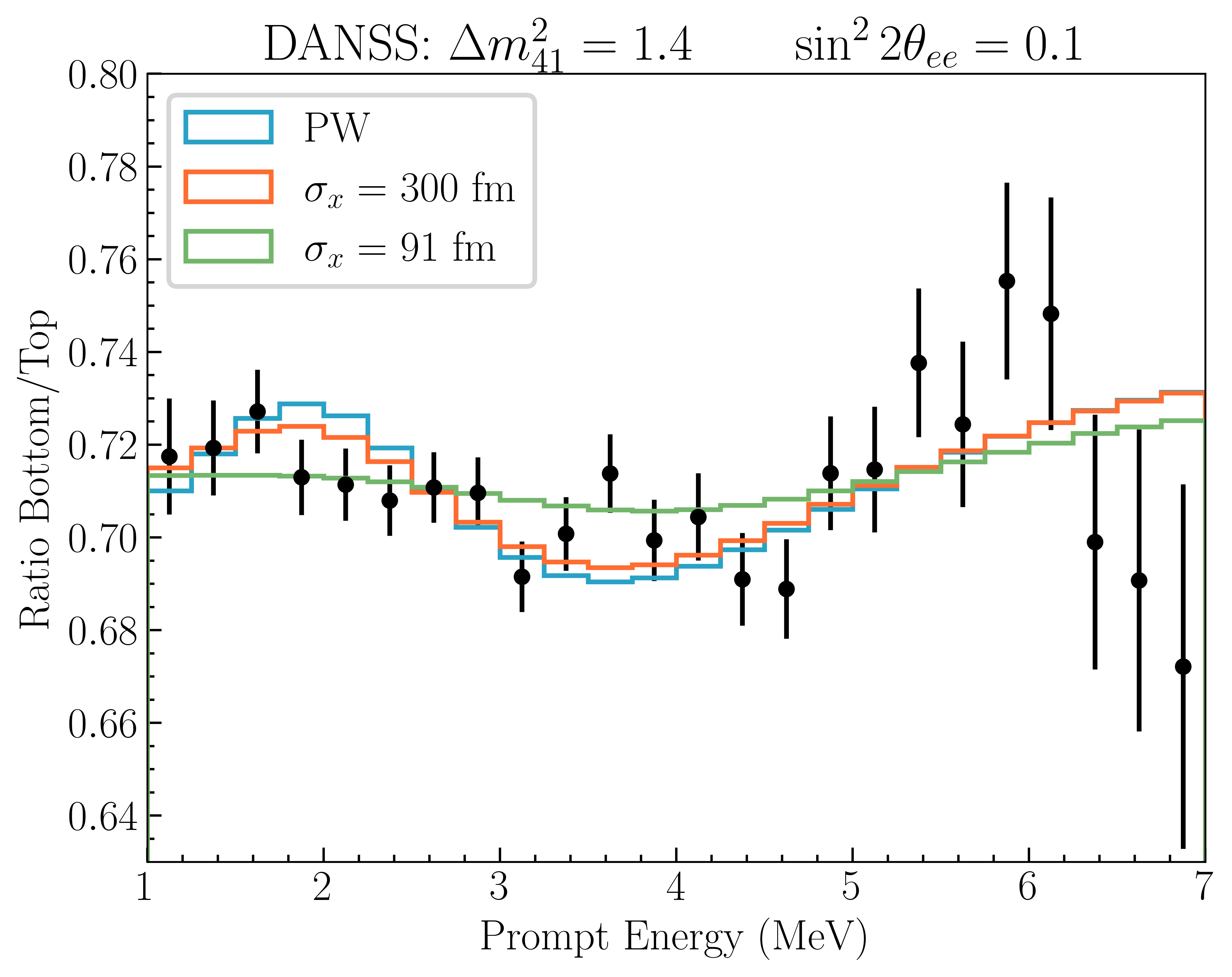}
\centering
\caption{The NEOS/RENO data ratio (left) and the DANSS data bottom/top ratio (right). Fits are to 3+1 and wave packet models discussed in sec.~\ref{sec:results}.  Plane Wave (PW) is 3+1, while $\sigma_x < \infty$ is the wave packet model.\label{NEOSRenoDANSS}}
\end{figure}

The assumption has been that these features are due to unmodeled contributions to the reactor flux spectrum, in which case the features will have no $L$ dependence.
In response to this, reactor experiments have switched to methods aimed at reducing sensitivity to non-$L$-dependent structure.   Two approaches are shown in Fig.~\ref{NEOSRenoDANSS}.
The left plot attempts to remove non-oscillation structure in the NEOS~\cite{NEOS} experiment data through normalizing to the RENO data from the same reactor complex~\cite{RENO}.
The right plot presents data from DANSS, which moves the detector to regularly alternate data-taking at three locations from the reactor core, 10.7~m (top), 11.7~m , and 12.7~m (bottom).
This case shows the bottom-to-top ratio~\cite{DANSS}.
In principle, all remaining structure should be due to $L$-dependent new physics effects.

\begin{figure}[tb!]
 \includegraphics[width=0.7\textwidth]{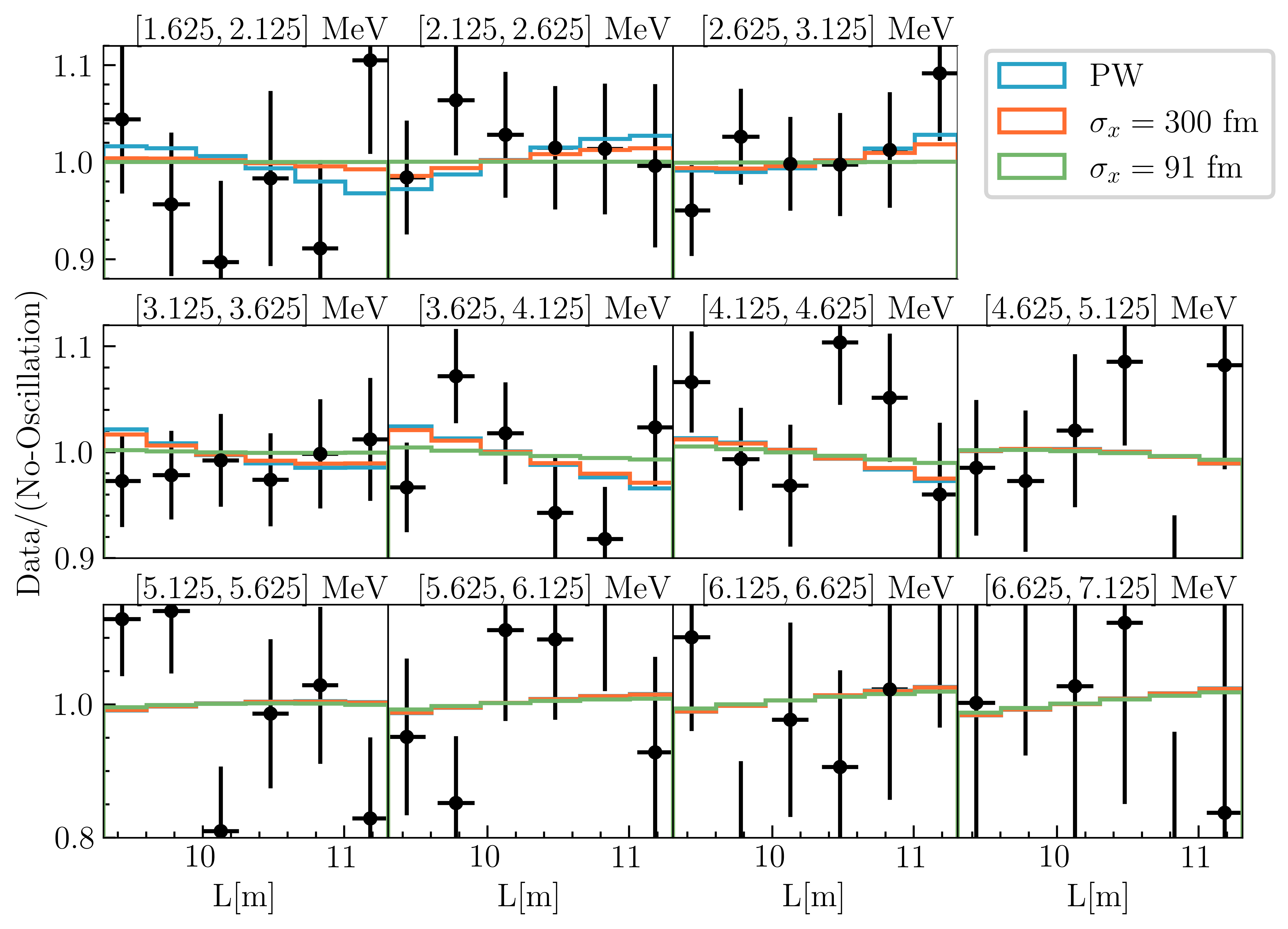}
\centering
\caption{The STEREO $L$ dependence in each energy bin compared to the same models presented for NEOS/RENO and DANSS.  Note the varying $y$-axis range.  Plane Wave (PW) is 3+1, while $\sigma_x < \infty$ is the wave packet model.\label{fig:STEREO}}
\end{figure}


\begin{figure}[tb!]
  \includegraphics[width=0.7\textwidth]{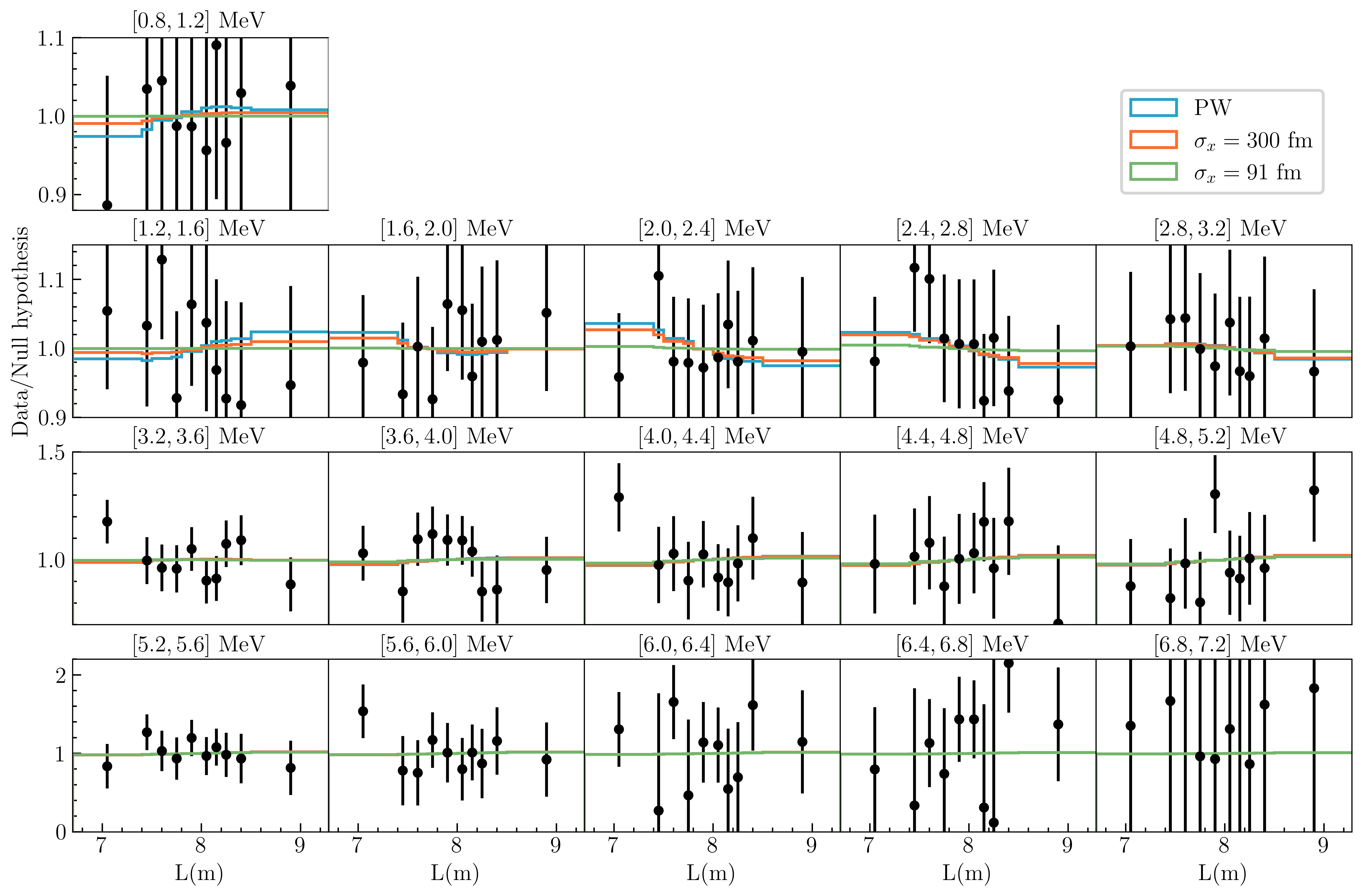}
\centering
\caption{PROSPECT $L$ dependence in each energy bin compared to the same models presented for NEOS/RENO and DANSS.  Note the varying $y$-axis range.  Plane Wave (PW) is 3+1, while $\sigma_x < \infty$ is the wave packet model.\label{fig:PROSPECT}}
\end{figure}

Comparing the structure seen in DANSS and NEOS/RENO in Fig.~\ref{NEOSRenoDANSS} to the predictions of Fig.~\ref{fig:damplo}, one sees that the extreme wavepacket and decay scenarios have very similar features to the structure that remains in these plots.
In Fig.~\ref{NEOSRenoDANSS}, we present the 3+1 plane wave (PW) and the two wave-packet examples considered earlier so that the effect is clear.
The concern is that, in practice, non-oscillation excess and deficits did not perfectly cancel due to systematic effects.
Considering DANSS as an example, the peak at 6~MeV may be from imperfect cancellation of the 5~MeV bump leaving a tail.  
The dip at higher energy may be due to imperfect cancellation of the high energy dip seen in Fig.~\ref{fig:mostreactors}.
It is difficult for those doing global fits to conjecture why there would not be full cancellation because we do not have hypotheses for the source of the underlying structure.
Therefore, it is very important that those who perform the reactor experiments revisit the potential systematic uncertainties in the reactor fits to understand if this is signal or an uncharacterized background.

Regardless of unexpected energy dependence, if one considers the $L$ dependence in each energy bin, then, in principle, an oscillation signature should be visible.   The STEREO and PROSPECT data are treated in this manner, as per the recommendation of those collaborations.    STEREO data, as seen in Fig.~\ref{fig:STEREO}, are fit for the shape of the $L$ dependence in individual energy bins, with the overall rate in each energy bin left as a free parameter. In the case of PROSPECT, the prediction at each baseline and energy is scaled according to the ratio of the total observed and predicted event rate across all baselines for each energy bin. Here, to illustrate this, in Fig.~\ref{fig:PROSPECT} we show the ratio of the data scaled to the prediction. In both cases, the effects we are considering in this discussion, indicated by the blue and orange lines, are small compared to the data uncertainty, and so these data are in agreement with the more precise DANSS and NEOS/RENO conclusions.

Our global fits rely on the $L$ and $E$ (i.e. ``shape'') dependence of the reactor data and not on the normalization for the following reason.   The first-principles prediction for the reactor event rate normalization has been observed to be higher than the measured data, an effect called the Reactor Antineutrino Anomaly or RAA.  The observation was first reported in 2011~\cite{Mention:2011rk}, and since that time, several groups have worked to improve the reactor predictions to address the issue.  The results are summarized in Ref.~\cite{Giunti:2021kab}, and range in the ratio of data-to-prediction from $0.925_{-0.023}^{+0.025}$ for the HKSS model to $0.975_{-0.021}^{+0.022}$ for the KI model.  Because of this large theoretical uncertainty, we rely on ``shape'' in our global fits.

\section{About the Fitting Code and Methods Used to Quantify Results \label{sec:quality}}

The fitting code uses a Markov Chain Monte-Carlo (MCMC) as described in Ref.~\cite{Diaz:2019fwt}.
Each experiment is considered independently, but individual experimental systematics are included based on information from the relevant collaboration.
The MCMC has the convenient property of sampling the model space, which allows us to use its sample points to display confidence regions and exclusion curves.
The confidence regions are described in two dimensions ($\Delta m^2$ and $\sin^2 2\theta$) using Wilks' theorem with 2 degrees of freedom on the profiled $\chi^2$, and this is what we report here.
We have compared the Wilks' theorem approach to a Bayesian approach in the past and find agreement, with more allowed space in the Bayesian case~\cite{Diaz:2019fwt}, but do not present those results here.  

Specifically, to quantify the quality of the fits, we use $\Delta \chi^2= \chi^2_{null} - \chi^2_{model}$ unless otherwise noted.
The $\Delta \chi^2$ subtracts the contribution from bins irrelevant to the model.

The focus of concern over many years has been the compatibility of subsets of data within the overall collection of global data.
For a detailed discussion see Ref.~\cite{Diaz:2019fwt}, but in particular, the appearance (app) (column 1 of Table~\ref{table:data}) and disappearance (dis) sets (columns 2 and 3 of Table~\ref{table:data}) when fit separately in a 3+1 model, show little overlap in $\Delta m^2$ of preferred parameter space.
The concern is that this points to different sources of anomalous effects in the two subsets, even in the presence of a large improvement in the global fit indicated by the $\Delta \chi^2$.  

The agreement within subsets is traditionally measured through the parameter goodness of fit (PG) test~\cite{Maltoni:2003cu}.
Along with the global fit (glob), the two subsets are fit separately allowing one to form an effective $\chi^2$:
\begin{equation}
\chi^2_{PG}=\chi^2_{glob}-(\chi^2_{app}+\chi^2_{dis}).  \label{chi2pg}
\end{equation}  
The number of degrees of freedom is then:
\begin{equation}
N_{PG}=(N_{app}+N_{dis}) - N_{glob},\label{npg}  
\end{equation}
where each $N$ is the number of independent parameters involved in the
given fit.
To be specific, for a model important to this discussion, for the 3+1+WP case, glob and dis will each have four parameters, $\Delta m^2$, $|U_{e4}|^2$, $|U_{\mu 4}|^2$ and $\sigma_x$, and app will have two parameters, $\Delta m^2$ and $|U_{e4}|^2 |U_{\mu 4}|^2$, so $N_{PG}=(4+2)-4=2$.
In contrast, in the example of the 3+1+dk case, where $\Gamma$ is fit in all data subsets, $N_{PG}=(4+3)-4=3$.
The probability is based on assuming a $\chi^2$ distribution for $\chi^2_{PG}$ and $N_{PG}$ and is defined as the ``tension'' between the subsets.  

\section{Results \label{sec:results}}

It has been clear for many years that 3+1 is an inadequate model to explain the data.
However, as we present results, we begin with this model for three reasons.
First, it is widely used, even if known to be inadequate.
Second, the features that appear are instructive.
Third, the comparison to models with damping (3+1+WP and 3+1+dk) will highlight that those models are better at describing the data.
For completeness, we also include results from 3+2 and 3+3 to show that those models result in less improvement.
That points to a preference for less structure (damping) in the data than for more structure (many mass splittings).
The results are summarized in Table~\ref{table:fitquality}.

\begin{table*}
\begin{centering}
\begin{tabular}{l | c | c | c | c | c }
Fit type:		    &	3+1	&	3+2	&   3+3 &   3+1+WP &	3+1+dk	 \\ \hline
(Null vs Sterile)		&		&		&       &       &		\\
	$\Delta \chi^2$	&	46.5 &	56.9 &	67.4 &  61.1 &	60.6 \\
	$\Delta dof$	&	3	&	7	&	12 &    4 & 4	\\
	$p$-value	&	4.4E-10	&	6.3E-10	&	9.8E-10	&   1.7E-12 &   2.2E-12\\
	$N\sigma$	&	6.2	&	6.2	&	6.1 &   7.1 &  	7.0 \\ \hline
(3+1 vs Other)		&		&		&		&       &\\
	$\Delta \chi^2$	&		&	10.1*	&	20.6* & 14.6 &  14.1	\\
	$\Delta dof$	&		&	4	&	9 & 1 & 1 \\
	$p$-value	&		&	3.9E-2	&	1.5E-2 & 1.3E-4 &  1.7E-4	\\
	$N\sigma$	&		&	2.1	&	2.4 & 3.8 & 3.8 \\ \hline
(PG Test)		&		&		&		&       &\\
	$\chi^2_{app}$	&	148.2	&	131.8	&   131.1& 148.2    & 146.8\\
	$N_{app}$	    &	2	    &	5	    &	9    & 2	    & 3\\
	$\chi^2_{dis}$	&	639.8	&	626.1	&	625.2& 638.4	    & 641.4\\
	$N_{dis}$	    &	3	    &	6	    &	9    & 4	    & 4\\
	$\chi^2_{glob}$	&	816.1	&	796.2    &	785.7& 801.5	& 807.5\\
	$N_{glob}$	    &	3	    &	7	    &	12   & 4        & 4\\
	$\chi^2_{PG}$	&	28.1	&	38.4    &	29.5 & 14.9     & 19.3\\
	$N_{PG}$	    &	2	    &	4	    &	6	 & 2        & 3\\
	$p$-value	    &	7.9E-07	&	9.3E-08 &	4.9E-05& 5.8E-04  & 2.4E-4\\
	$N\sigma$	    &	4.9	    &	5.3	    &	4.1	 &3.4       & 3.7\\ \hline
\end{tabular}
\caption{\label{table:summary} A summary of the quality of the fits. Columns correspond to the five types of fits. Top section: Comparison of quality of null to each fit including sterile neutrinos; Second section: Comparison of 3+1 to the extended models; Bottom section: PG test results for each model, where Eqs.~\ref{chi2pg} and \ref{npg} explain how $\chi^2_{PG}$ and $N_{PG}$ are determined.  The asterisk (*) denotes that the 3+2 and 3+3 models are compared to a 3+1 fit that does not include IceCube, as explained in the text.}
\label{table:fitquality}
\end{centering}
\end{table*}

\begin{figure}[tb!]
\centering
 \includegraphics[width=0.45\textwidth]{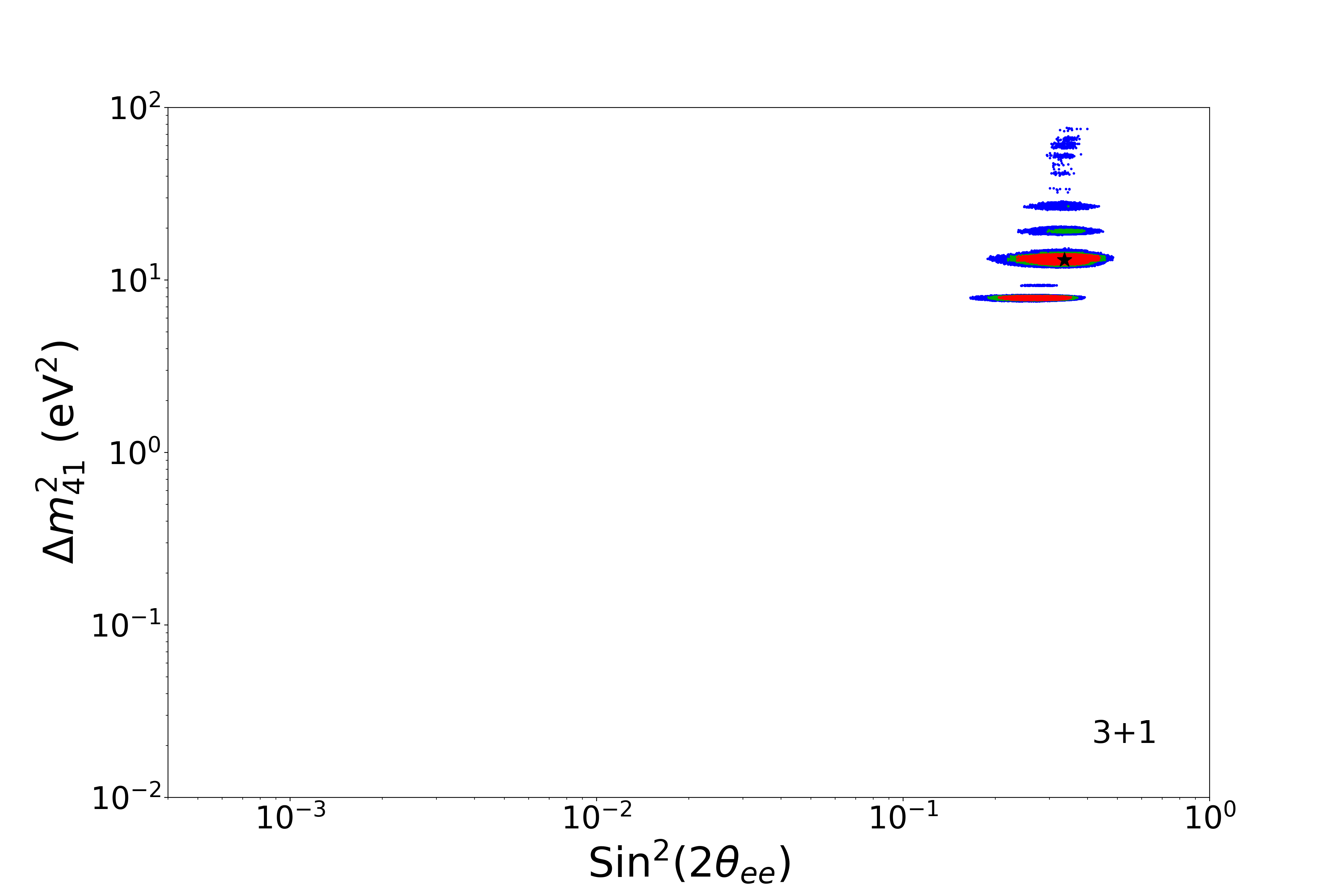}
 \includegraphics[width=0.45\textwidth]{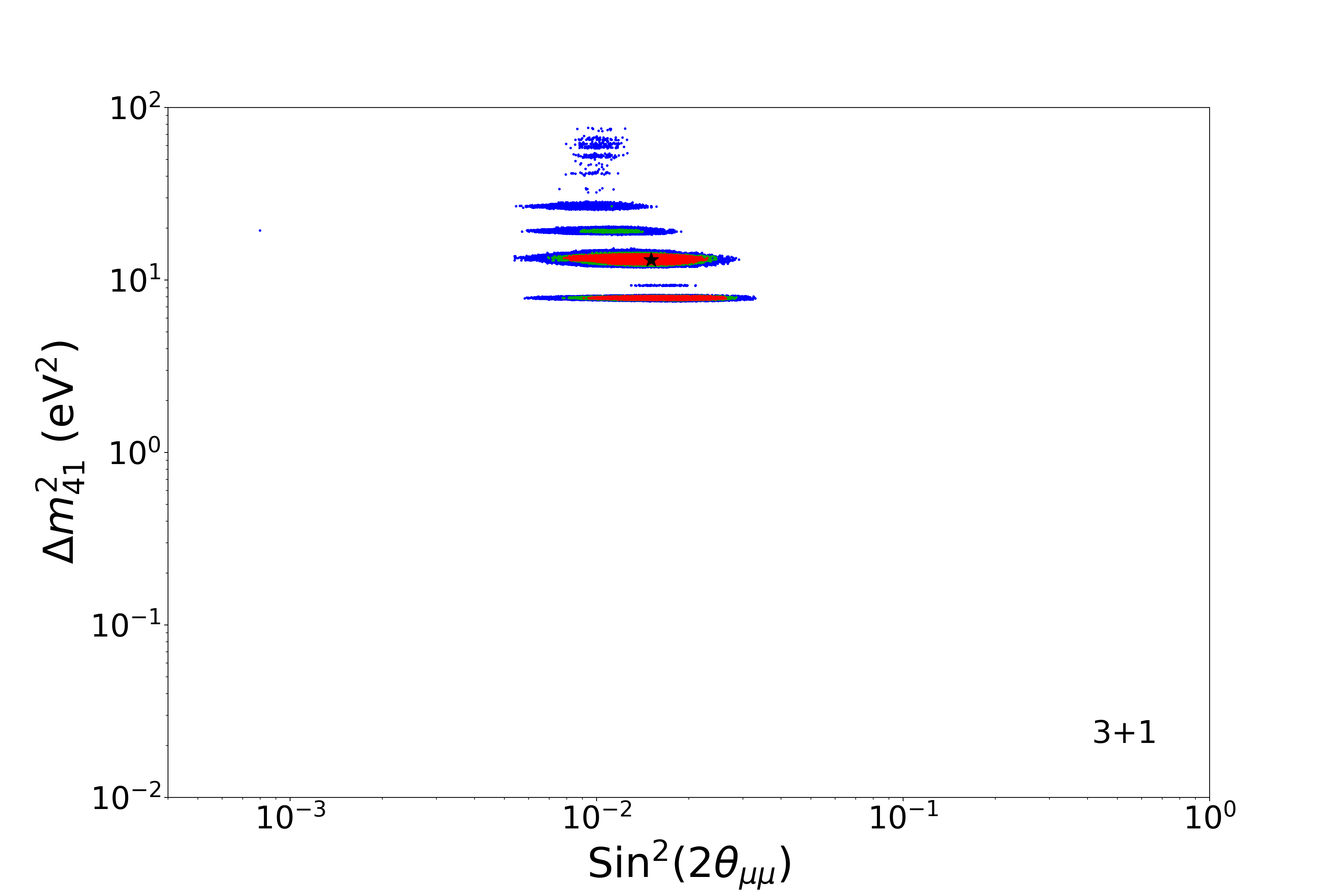}
 \includegraphics[width=0.45\textwidth]{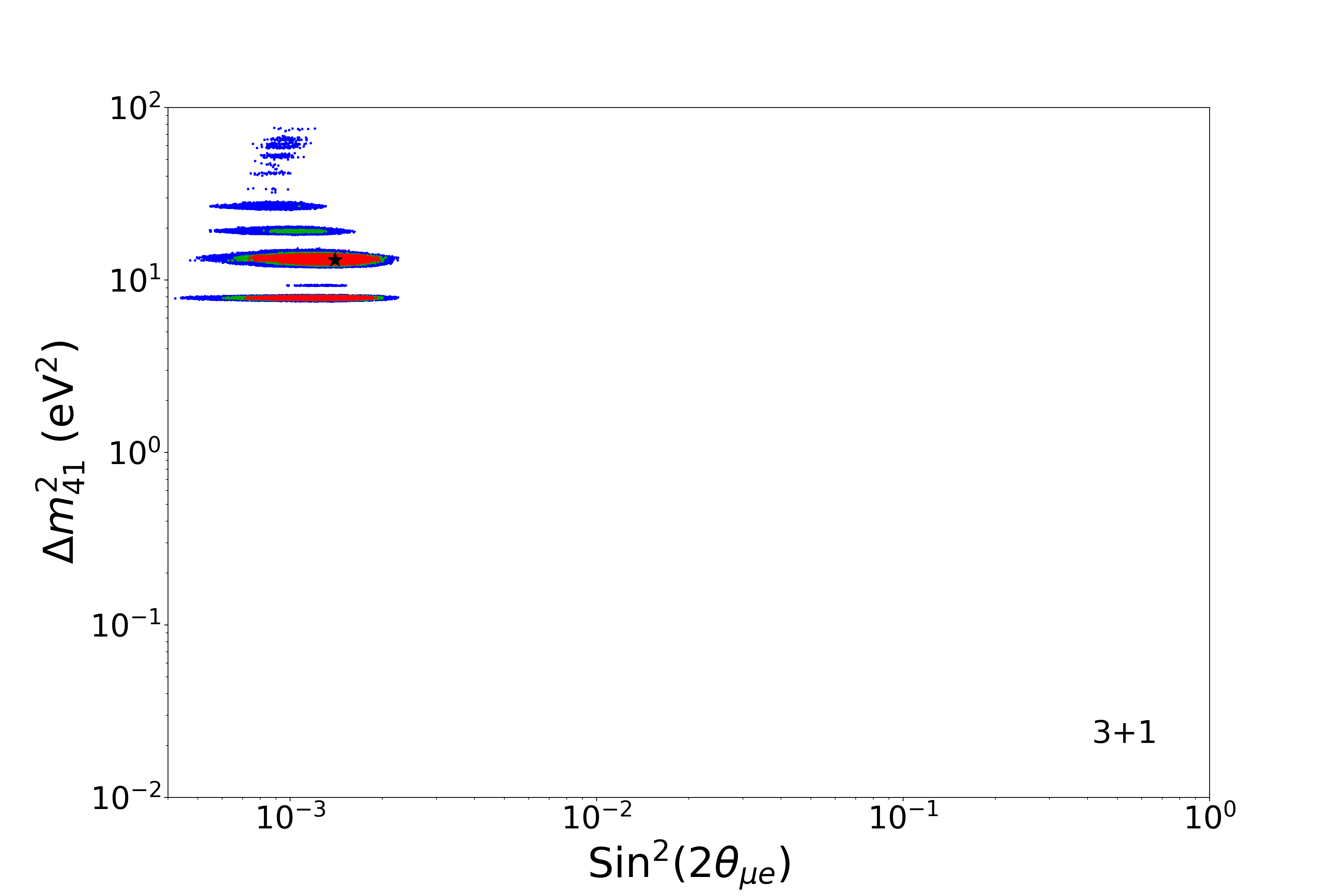}
\caption{\label{fig:2022} The plane wave 3+1 fit results to the 2022 data sets.  Top Left:  electron-flavor disappearance; Top Right: muon-flavor disappearance; Bottom: appearance.   Note that each plot is a projection profiled from the 3D fit space.}
\end{figure}

\subsection{3+1 (Plane Wave) Fit results}

This section describes the 2022 results for fits to the commonly used, plane-wave 3+1 model.
Results are shown in Fig.~\ref{fig:2022}.
One sees that the new fits prefer a $\Delta m^2=13$~eV$^2$ compared to 1.3~eV$^2$ from the 2019 fits---an order of magnitude shift.
This large shift is due to the interplay of the new inclusion of BEST, which has little $L$ dependence, and so fits well at large $\Delta m^2$ but requires very large mixing, and the updated collection of reactor data that is poorly fit by a 3+1 model, but that does not produce a strong limit above $\Delta m^2=10$ eV$^2$.
This forces the fit point to just above the reactor limit in the $\nu_e$ disappearance samples, as can be inferred from Fig.~\ref{fig:2022}, top left, and the other samples can find solutions in the same range, making this the best-fit point.
The best-fit mixing matrix parameters are $|U_{e4}|^2=8.5 \times 10^{-2}$ and $|U_{\mu4}|^2=3.8\times 10^{-3}$.
The $\Delta \chi^2 /$DOF for 3+1 compared to null is 46.5/3, which is an extraordinary improvement from adding one sterile neutrino---far more than 5$\sigma$.
This says that some effect that is not in the null prediction, and that has features similar to 3+1, is very strongly preferred by the data.

The range of the axes in the plots of Fig.~\ref{fig:2022} shows the parameter space searched for this analysis.  The plots in Fig.~\ref{fig:2022} are projections profiled from the 3D parameter space into the 2D planes.    Although a large space is searched, the allowed regions are small. Since there is only one answer, that, in itself, is not an issue.  The best fit $\chi^2$ minimum is very deep compared to null.

\begin{figure}[tb!]
 \includegraphics[width=0.5\textwidth]{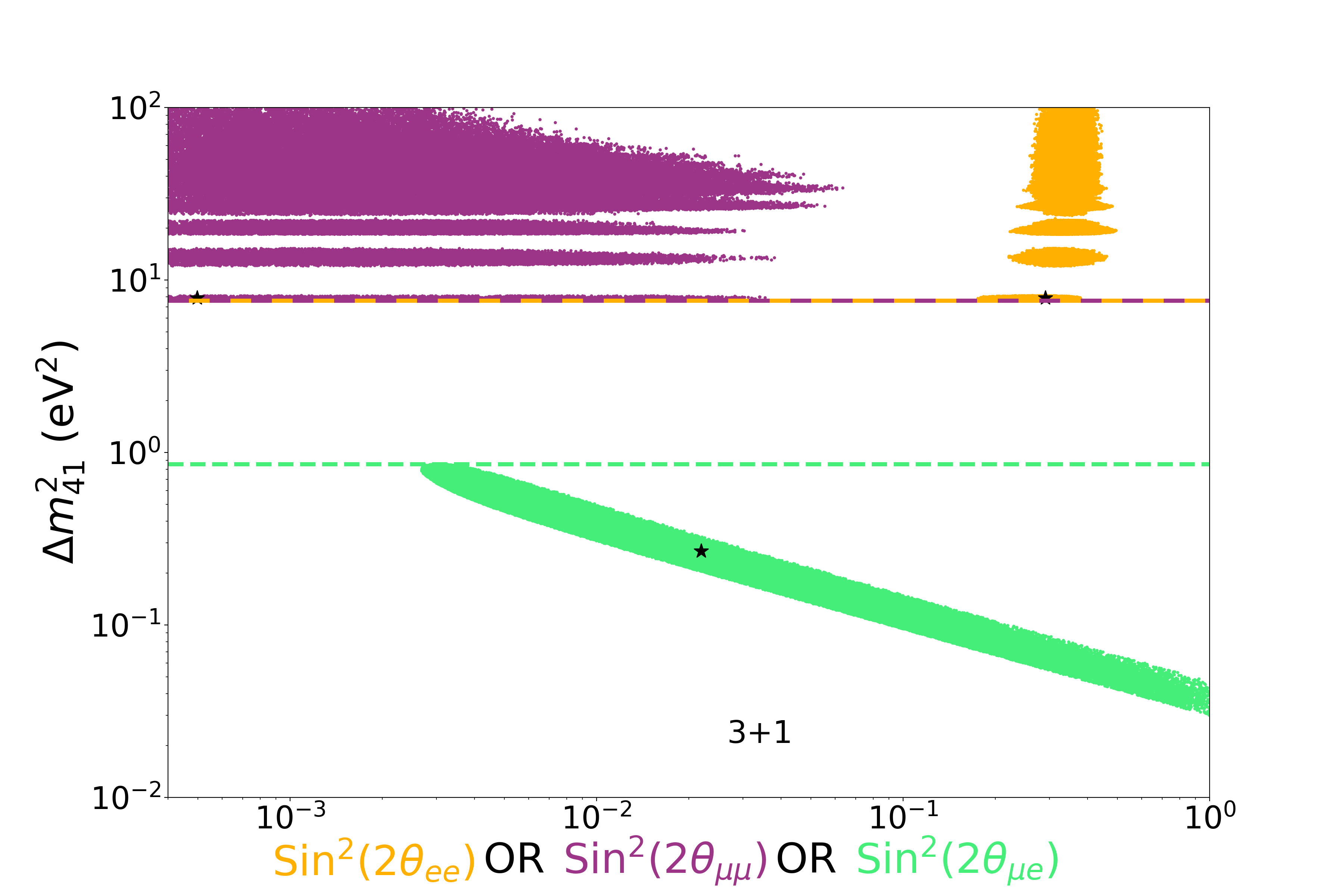}
\centering
\caption{\label{fig:tension} Illustration of the tension within the 3+1 fit.
Results of separate app and dis fits for the $\nu_\mu$ disappearance (populating upper left),  $\nu_e$ disappearance (populating upper right), and appearance (populating lower region) data sets.   The 95\% region, equivalent to the green in Fig.~\ref{fig:2022}, is shown in a different color for each of the mixings. Dashed lines guide the eye on the level of the gap in preferred $\Delta m^2$.}
\end{figure}

With that said, there is well-known tension internal to the data sets that has become worse with these 2022 fits.
In Fig.~\ref{fig:tension}, we show the separate 95\% CL allowed
regions for electron and muon disappearance (dis), orange and purple respectively, and muon to electron appearance (app) overlaid on the same plot in green. 
One sees that the $\nu_\mu$ disappearance solutions populate the upper left (high $\Delta m^2$, small mixing angles) and $\nu_e$ disappearance populates the upper right (high $\Delta m^2$, large mixing angles).
Since $\sin^2 2\theta_{ee}$ and $\sin^2 2\theta_{\mu\mu}$ are independent, these two data sets are compatible with a best combined fit of $\Delta m^2=7.8$~eV$^2$.   As discussed in Sec.~\ref{sec:models}, the underlying matrix elements for $\sin^2 2\theta_{ee}$ and $\sin^2 2\theta_{\mu\mu}$ combine to make $\sin^2 2\theta_{\mu e}$, however this can be accommodated in the fits.   The tension primarily arises from the large gap in preferred $\Delta m^2$ between the disappearance data sets and the appearance data set.  The appearance data set has a preferred fit at $\Delta m^2=$0.24~eV$^2$.   To emphasize this gap, we show dashed lines consistent with the lowest $\Delta m^2$ allowed solutions for $\nu_e$ and $\nu_\mu$ data in orange and purple, and the highest allowed solution for appearance in green.   This means that the app and dis tension is high, and that the global fit $\Delta m^2$ at 13~eV$^2$ is a poor compromise for both.
Making this quantitative using the PG test discussed in Sec.~\ref{sec:quality}, the level of disagreement is $\chi^2_{PG}/N_{PG}= 28.1/2$, which represents a 4.9$\sigma$ tension.

As with the cosmological data, we have not included solar neutrino data in our fits because the interpretation of these data is model-dependent (see for example, Ref.~\cite{Davoudiasl:2023uiq}).  However, we have tested the effect of solar neutrino data on the studies by introducing a constraint on $\sin^2 2\theta_{ee}$ from Ref.~\cite{Goldhagen_2022}.  The result changes the $\Delta \chi^2$/dof from 46.5/3 to 38.2/3.  Thus, 3+1 remains a substantially better solution than the three neutrino solution.

\begin{table}[t]
\begin{center}
\begin{tabular}{|l|c|c|}  \hline
Fit & App vs. Dis & $\Delta m^2$ best fit (eV$^2$)\\
 & tension & App., Dis, Global \\ \hline
3+1 &  4.9$\sigma$  & 0.24, 7.8, 13 \\
3+1+WP & 3.5$\sigma$  & 0.24, 1.4, 1.4\\
3+1+WP, No MiniBooNE & 2.1$\sigma$ & 0.84, 1.4, 1.4\\
 \hline
\end{tabular}
\caption{\label{table:wpres}  Progression of reduction of tension of fits from 3+1 to 3+1+WP to 3+1+WP with MiniBooNE removed from fit.}
\end{center}
\end{table}

\subsection{3+1+WP Fit Results}

\begin{figure}[tb!]
  \includegraphics[width=0.5\textwidth]{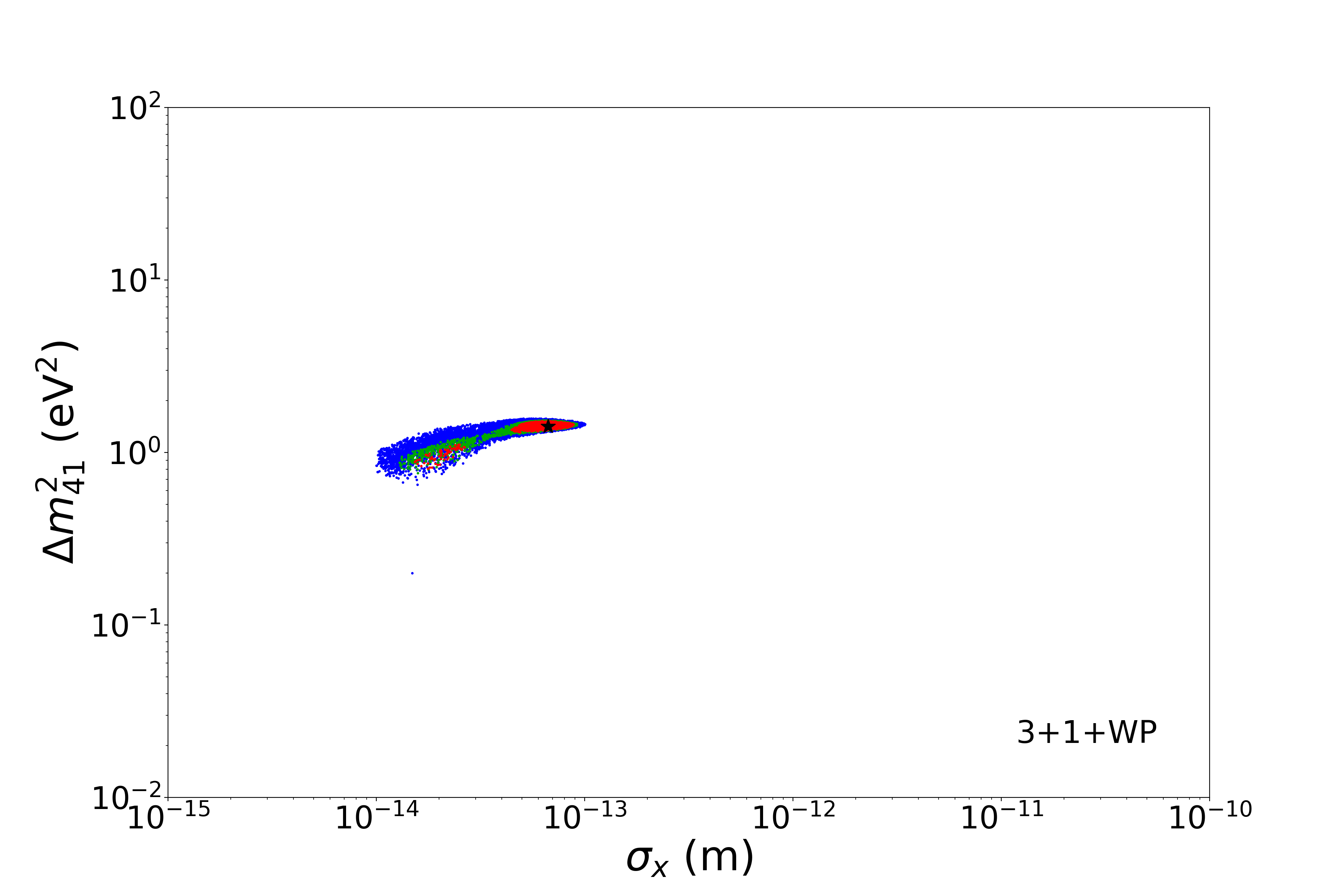}
\centering
\caption{\label{fig:2022PW} The relationship between the fit mass splitting and the characteristic wavepacket size, $\sigma_x$, in the 3+1 global fit.   Note that this plot is a projection profiled from the 4D fit-space.}
\end{figure}

Introducing 3+1+WP leads to a substantial improvement in the fit results.
We have qualitatively discussed the reason for this in Secs.~\ref{sec:WP} and \ref{sec:reactdiscuss}. 
Here, we discuss the quantitative results of the global fit, and then consider the implication of the results from medium- and long-baseline reactor experiments in a simplified model.

\subsubsection{Results of the Global Fit}
 
The improvement in the fit compared to the null is $\Delta \chi^2$/dof = 61.1/4. The improvement is 14.6/1 compared to 3+1.
The best-fit $\Delta m^2=1.4$ eV$^2$ which is in excellent agreement with the 2019 fit.  This $\Delta m^2$ shift from the higher 3+1 solution arises because the 3+1+WP result damps and weakens the predicted low energy oscillations for reactor experiments.
The best-fit mixing matrix parameters are $|U_{e4}|^2=8.8 \times 10^{-2}$ and $|U_{\mu 4}|^2=3.7\times 10^{-3}$.   Plots of the allowed regions are provided in Appendix~\ref{AppWP}.
For the wavepacket size, the best fit finds $\sigma_x=67$~fm.   
Fig.~\ref{fig:2022PW} shows the $\sigma_x$ fit as a function of $\Delta m^2$.  
This can be compared to the reactor data fit alone for the wavepacket effect, which prefers 91 fm corresponding to the green curve on Fig.~\ref{fig:damplo}.   Thus, this model is introducing strong damping on the reactor prediction at low energy but the best fit allows shape effects at large energies.  

Next, let us consider the tension in this model.  The 3+1+WP model yields $\chi^2_{PG}$/$N_{PG}=14.9/2$.
This represents 3.4$\sigma$ tension, which is still significant but is substantially higher probability than the 4.9$\sigma$ tension of the 3+1 case.  We summarize the probabilities in Table~\ref{table:summary}.
Another way to say this is that the $\chi^2_{PG}$ improved by 13.2 units with the addition of only 1 degree of freedom.

\begin{figure}[tb!]
 \includegraphics[width=0.49\textwidth]{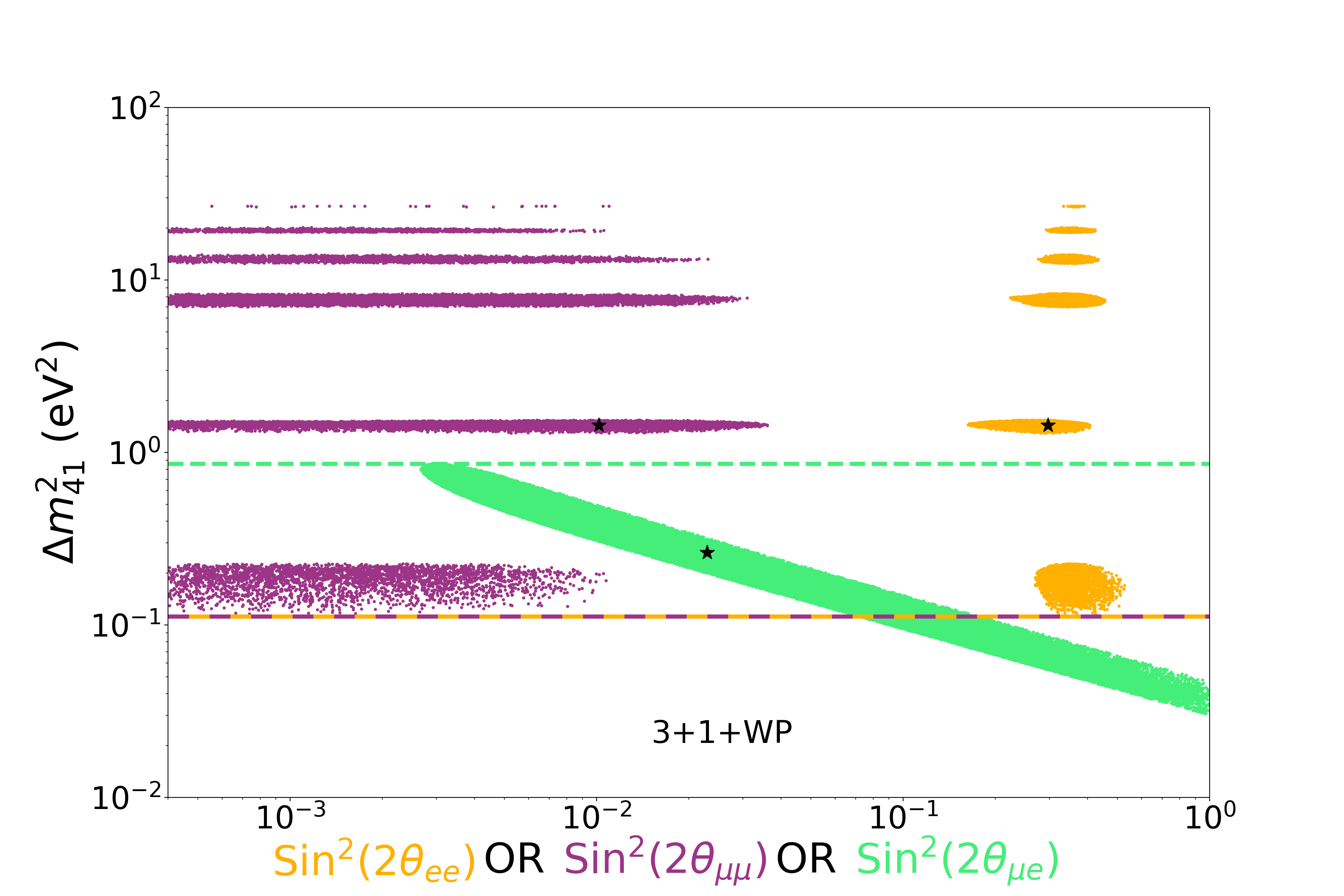}
 \includegraphics[width=0.49\textwidth]{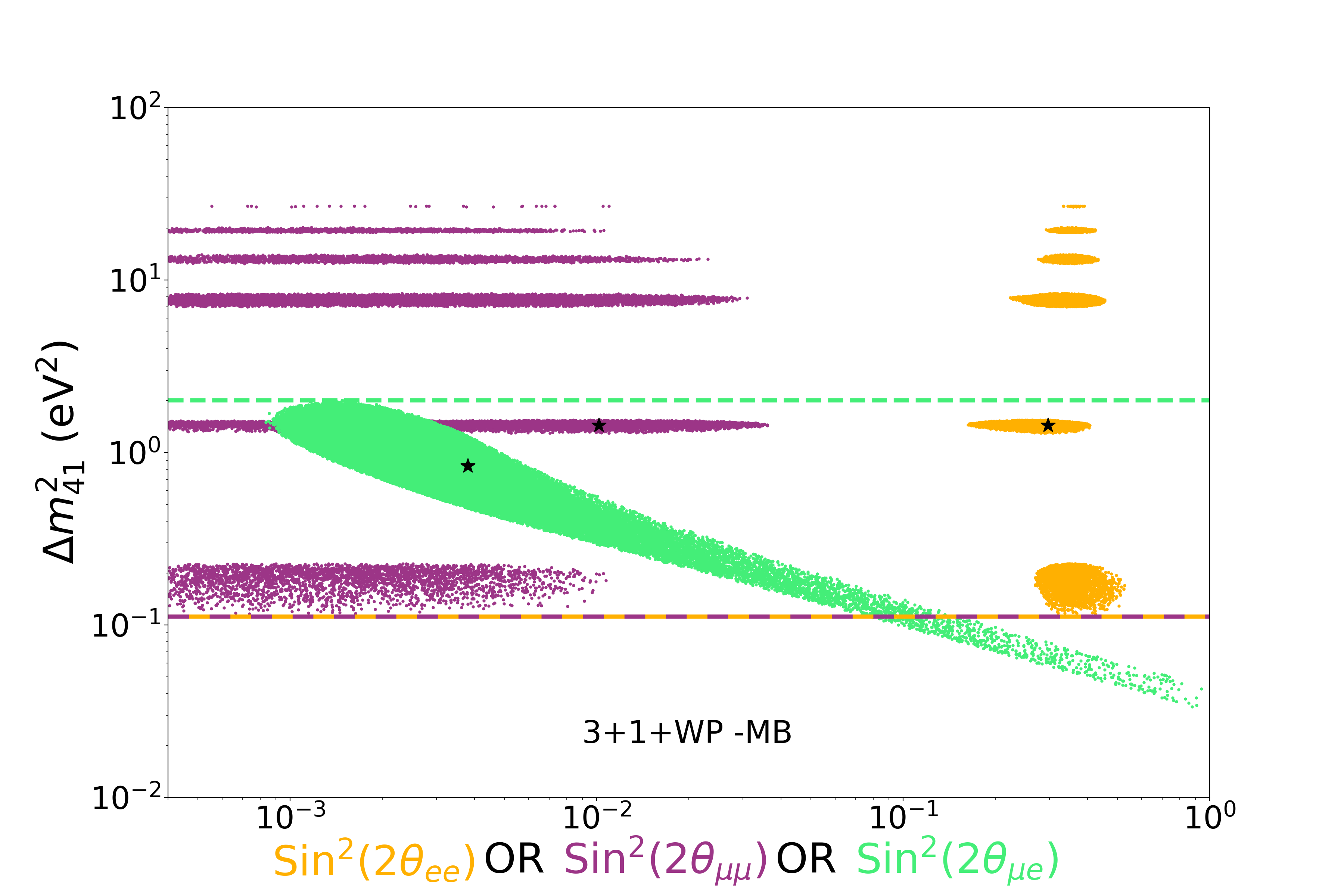}
\centering
\caption{\label{fig:tensionPW} Left:  Illustration of the tension within the 3+1+WP fit.  The gap between the dis and app allowed regions is greatly reduced.  Right:  Same but with the MiniBooNE appearance data set removed from the fits.   At this point, one sees overlap of the 2$\sigma$ regions between the app and dis allowed regions.}
\end{figure}

Fig.~\ref{fig:tensionPW}, left, allows one to visualize the improvement. 
Compared to Fig.~\ref{fig:tension} for 3+1,  dis and app allowed regions for 3+1+WP no longer show a gap and, instead, have substantial overlap.    However, one still sees a gap between the dis and app best fit points, and that indicates some tension remains. 

\begin{figure}[tb!]
 \includegraphics[width=0.5\textwidth]{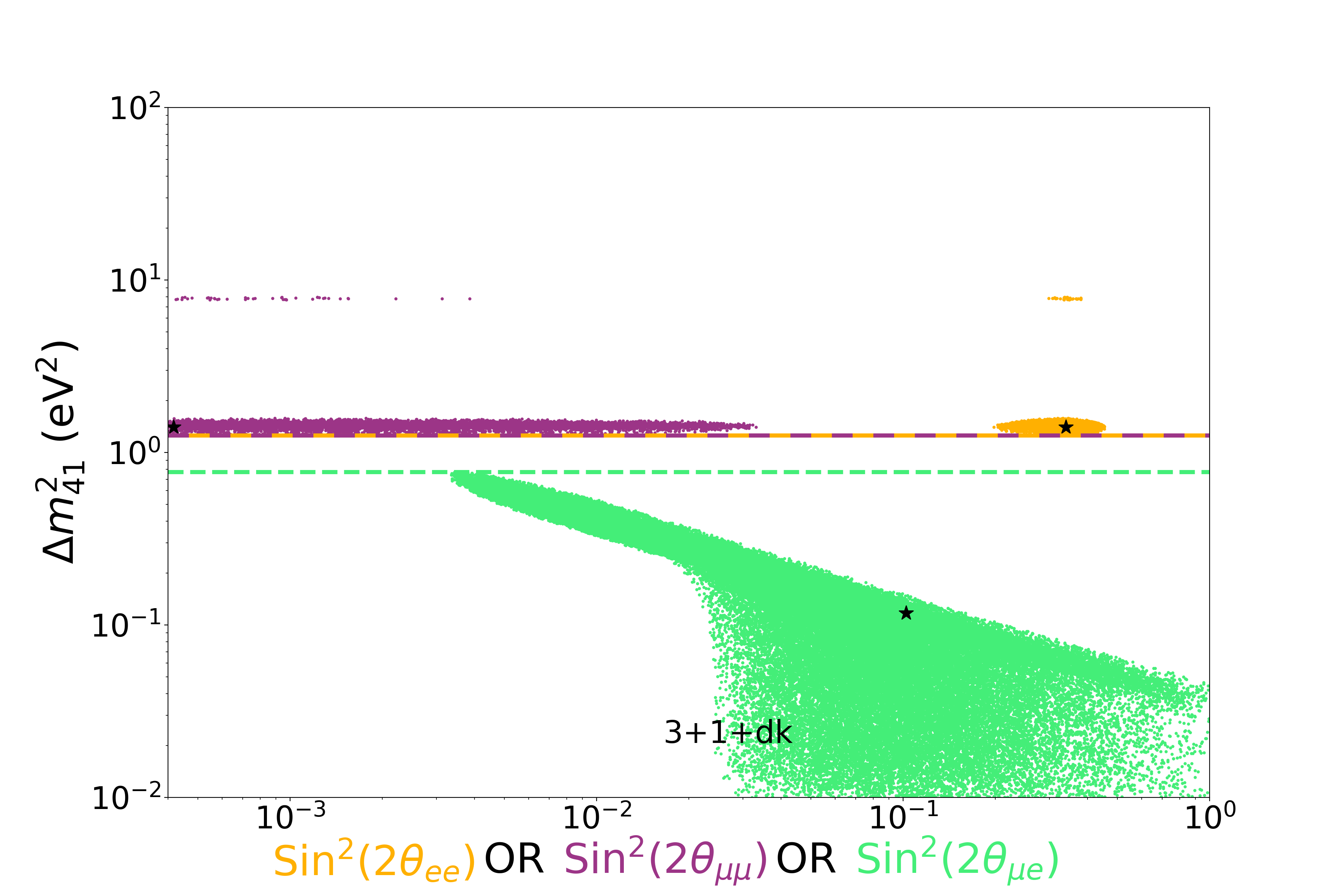}
\centering
\caption{\label{fig:tensiondk} Illustration of the tension within the 3+1+dk fit.  The gap between the dis and app allowed regions is greatly reduced relative to 3+1 (Fig.~\ref{fig:tension}), but not to the level of 3+1+WP (Fig.~\ref{fig:tensionPW}, left)}
\end{figure}

We can test for the primary source of the remaining tension in the model by removing each experiment while leaving all others in the fit.  The largest improvement comes from removal of MiniBooNE.  MiniBooNE is unique among the experiments in that it appears to have two signal sources,  one that is consistent with $\nu_e$ CCQE-like scattering and an additional signal consisting of forward scatters \cite{MiniBooNE:2020pnu}, as described in Sec.~\ref{sec:MB}.   The fit of the total signal to the CCQE-like prediction with oscillations prefers $(\sin^22\theta$, $\Delta m^2)$=(0.807,0.043 eV$^2$) \cite{MiniBooNE:2022emn}, far from the best fit parameters of the other experiments.	Thus, removing this experiment improves the tension. The progression of improvement in tension is presented in Table~\ref{table:wpres}, and is also illustrated by the improved proximity of the app and dis best fits in Fig.~\ref{fig:tensionPW}, right, when MiniBooNE is removed.  Quantitatively, removing MiniBooNE leaves $2.1\sigma$ of tension, where no single experiment dominates.  

\subsubsection{Comment on medium- and long-baseline reactor data}

\begin{figure}[tb!]
 \includegraphics[width=0.5\textwidth]{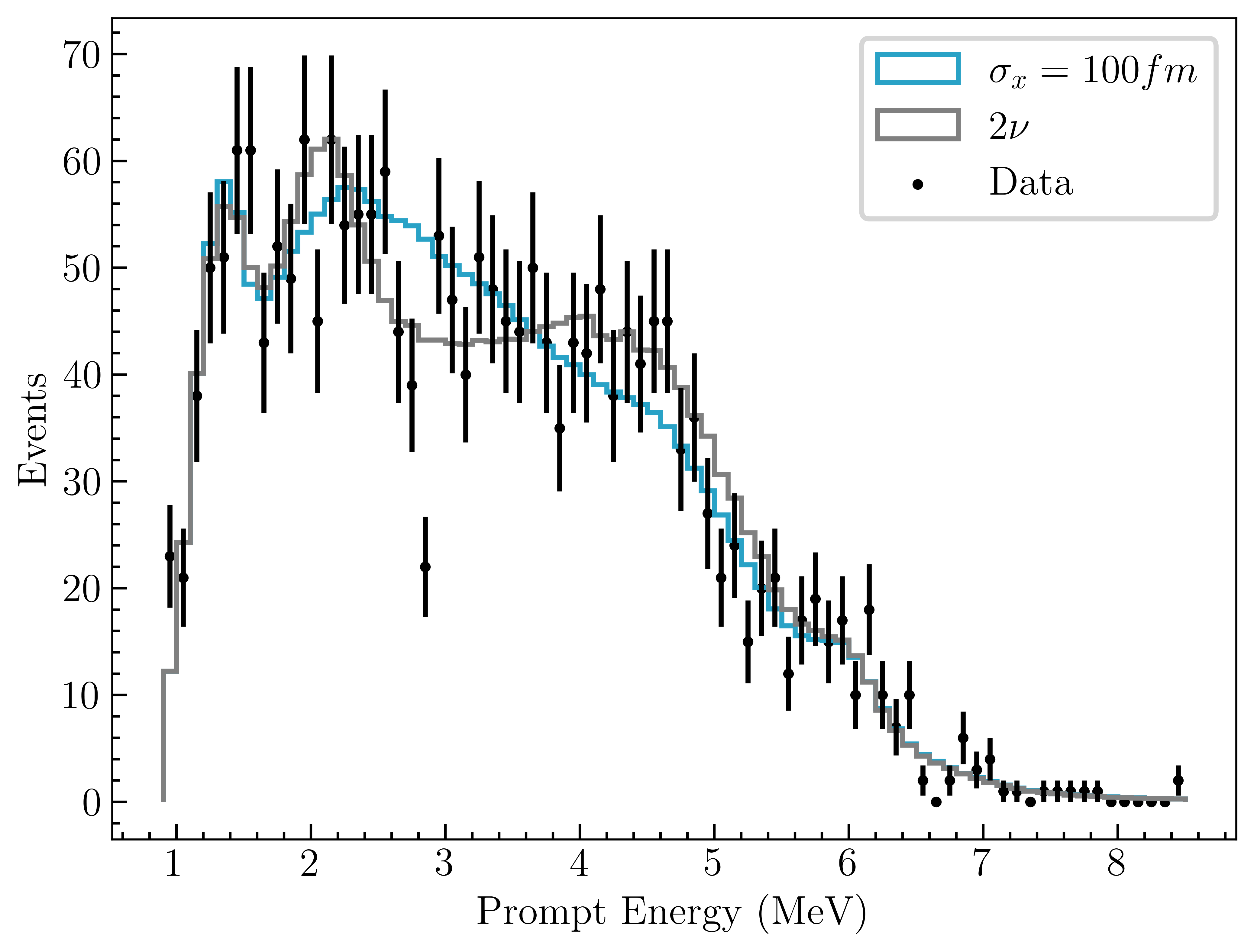}
\centering
\caption{\label{fig:KamLANDdata} Event count, including background, from the KamLAND data release.  The prediction for two-neutrino oscillations with $\Delta m^2=7.4\times 10^{-5}$ eV$^2$ is shown in gray.  The prediction also including a 100 fm wavepacket is shown in blue.}
\end{figure}

As discussed in Sec.~\ref{sec:models}, these global fits use the short-baseline, or ``sbl'' approximation.   Thus, the world’s data sets that are sensitive to low mass splittings, below $10^{-2}$ eV$^2$, are not included.   As a result, the data in our global fits does not include the medium- and long-baseline reactor data sets.    In a full global fit, that also includes the light neutrino parameters and the wavepacket effect, these data sets would influence the value of $\sigma_x$.   It is beyond the scope of this paper to expand to a full fit, however, here we consider simplified approaches to estimate the impact of these data sets. 
 
The Daya Bay and Reno medium-baseline reactor data were examined in a two neutrino oscillation model with wavepacket effect by de Gouv\^ea, et al., in Ref.~\cite{deGouvea:2020hfl}.   That group found that medium-baseline data prefer $\sigma_x > 95$ fm at 95\% CL.   This overlaps with our global fit that has an allowed region of $46<\sigma_x<110$ fm at 95\% CL.    Thus, the medium-baseline reactor data does not conflict with our result, and, in fact, a fit that goes beyond the sbl approximation to include the medium-baseline data is expected to narrow and strengthen the allowed region for $\sigma_x$.

Next, we explore the long-baseline reactor data.   At present, KamLAND supplies the only long-baseline reactor data, although JUNO will follow with additional results in the late 2020's \cite{Denton:2022een, Marzec:2022mcz}.    De Gouv\^ea,  et al., have shown in Ref.~\cite{deGouvea:2021uvg} that KamLAND imposes constrains $\sigma_x > 210$ fm at the 90\% confidence level in the three neutrino fit.  To explore the level that KamLAND disagrees with the 3+1+WP result, and to understand the source of the disagreement and its implications for JUNO, we have incorporated the KamLAND data as a constraint in our fits.

Fig.~\ref{fig:KamLANDdata} shows the data from the KamLAND data-release (including backgrounds)~\cite{PhysRevD.83.052002, kamland:data}.       The gray line overlays the result of the KamLAND-reported flux modified by a two-neutrino fit to the data that corrects for 5 MeV unexplained reactor-flux feature and fits the 76 bins with $>1$ event. The result prefers $\Delta m^2=7.4 \times 10^{-5}$ eV$^2$, which is very close to the KamLAND-reported best fit of $\Delta m^2=7.5 \times 10^{-5}$ eV$^2$~\cite{PhysRevD.83.052002}. The blue line is an example of the predicted oscillation including the wavepacket effect with $\sigma=100$ fm.  A striking feature of this plot is the low data point at prompt energy $E_{p}=2.85$ MeV.   This data point, which lies at a minimum of the two-oscillation (no wavepacket) prediction, represents a deviation from the fit with a Poisson probability of $2.9\times 10^{-4}$,  hence a very rare fluctuation.    The disagreement between this point and the wavepacket prediction is even worse, since that prediction (blue) lies above the two neutrino oscillation prediction (gray).

\begin{figure}[tb!]
\includegraphics[width=0.49\textwidth]{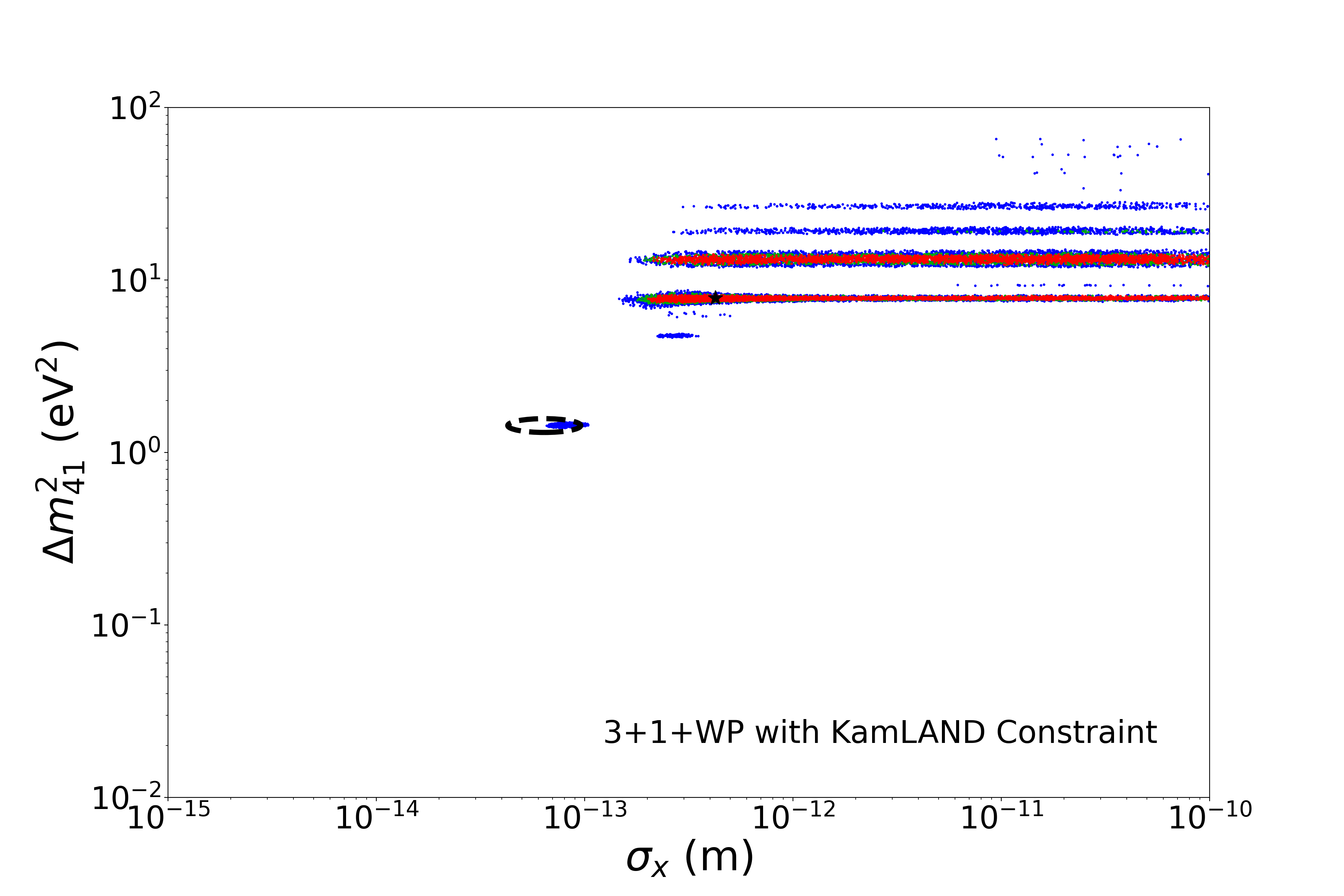} 
\includegraphics[width=0.49\textwidth]{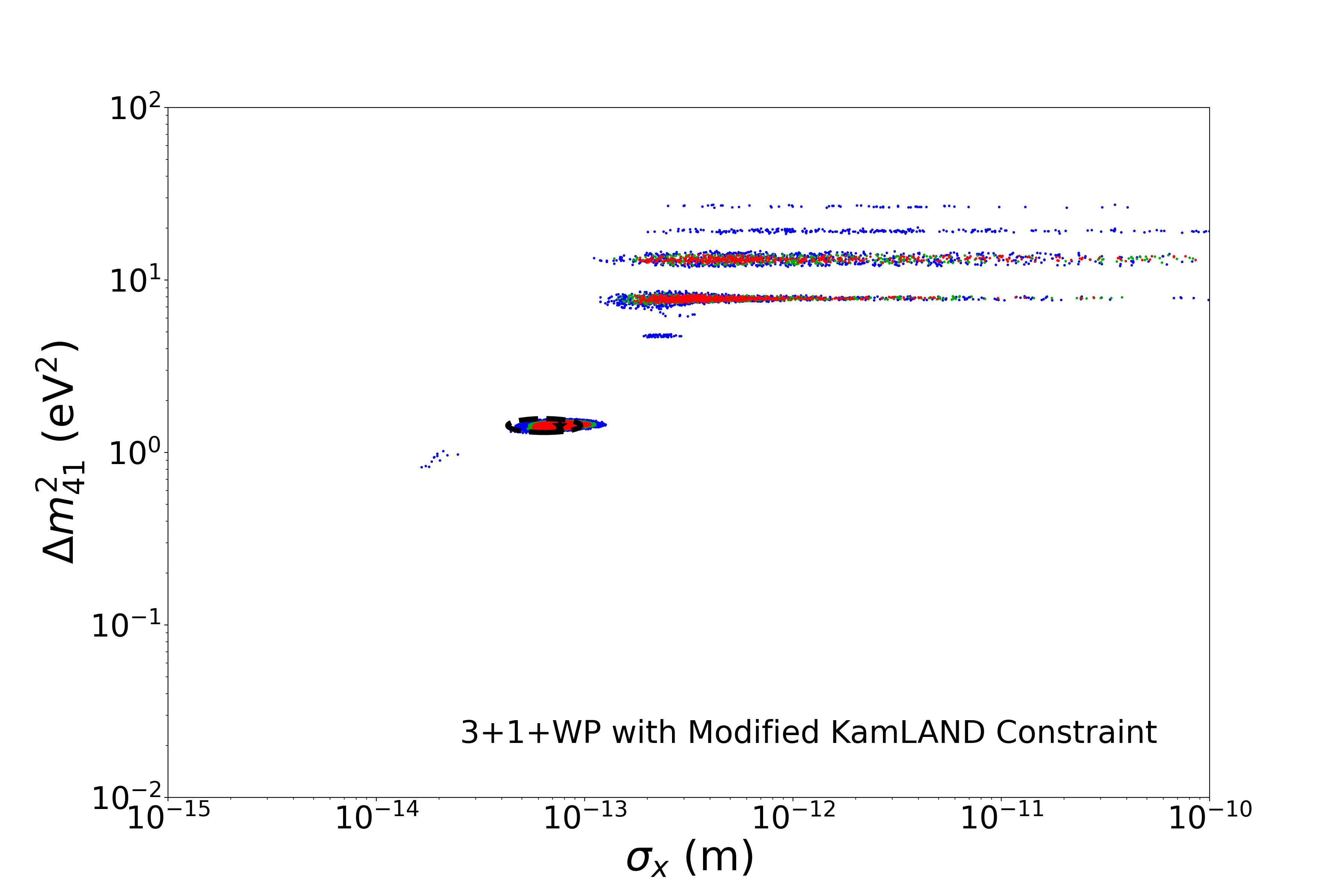} 

\centering
\caption{\label{fig:KamLANDWP} Two cases that bracket the expectation for JUNO for $\sigma_x$ (see text). Left: $\Delta m^2$ vs $\sigma_x$ in the 3+1+WP fit with full KamLAND constraint.  Right: The 3+1+WP case with the KamLAND constraint modified by excluding the low point at 2.85 MeV with probability of $2.9\times 10^{-4}$. The dashed enclosed region is the allowed region for 3+1+WP with no long-baseline reactor constraint (see Fig.~\ref{fig:2022PW}).}
\end{figure}

To explore the impact of the KamLAND data on the wavepacket discussion, we use the data as a constraint on our global fit, including statistical and systematic errors as described in the data release.   Fig.~\ref{fig:KamLANDWP} shows the $\sigma_x$ global fit result for the KamLAND data set.    For reference, the no-KamLAND global fit from Fig.~\ref{fig:2022PW} is indicated at 95\% CL by the dashed enclosed region.    The KamLAND data set prefers solutions for $\sigma_x$ at relatively high $\Delta m^2$, where the short baseline reactor results are smoothed by rapid oscillations.  However, a small island at $\Delta m^2=1.4$ eV$^2$ exists at 99\% CL that is consistent with the no-KamLAND fit.  	 The highly unlikely fluctuation at 2.85 MeV has a major impact on this result.    To see the effect, we fit with the 2.85 MeV point removed, where the best fit shifts to $\Delta m^2$ of 1.4 eV$^2$ and produces a closed contour with $52<\sigma<114$ fm at 95\% CL, as seen in Fig.~\ref{fig:KamLANDWP} (right).   Assuming the KamLAND point at 2.85 MeV is a statistical fluctuation, JUNO is most likely to observe a higher value.   Thus, the result of Fig.~\ref{fig:KamLANDWP} (left) and (right) can be expected to bracket the JUNO expectation, allowing $\sigma_x \approx 100$ fm. 

\subsection{3+1+dk Fit Results}

The 3+1+dk model yields a similar result to 3+1+WP, underlining the preference for a damping term.   
The best fit is $\Delta m^2=$ 1.4 eV$^2$; the best-fit mixing parameters are $|U_{e4}|^2=1.0 \times 10^{-1}$ and $|U_{\mu 4}|^2=5.5\times 10^{-3}$; and the best-fit $\Gamma=\SI{0.35}{\eV}$.  The example given in green in Fig.~\ref{fig:damplo}, right is from the reactor only fit at a similar value of $\Gamma=\SI{0.29}{\eV}$.   Plots of the allowed regions are provided in Appendix~\ref{Appdk}.
The difference between the models is that in this case the damping parameter affects all data sets rather than just the reactor experiments and has a slightly different dependence on $L/E$.
The overall goodness of fit matches 3+1+WP, with $\Delta \chi^2$ compared to 3+1 of 14.1 for only 1 additional fit parameter.
The tension in this model is compared to 3+1 and 3+1+WP in Table~\ref{table:fitquality}.
For this model, when fitting all experiments, the $\chi^2_{PG}$ is 19.3 for $N_{PG}=3$, hence the tension is at the $3.7\sigma$ level.
One can gain a visual sense of the relative improvement by comparing Fig.~\ref{fig:tensiondk} to Figs.~\ref{fig:tension} and \ref{fig:tensionPW}, left.
One sees that while there are differences in the allowed regions overall, 3+1+WP and 3+1+dk are improving the fits.

As a similar test to the wavepacket case, we have fit the reactor experiments separately.  We find $\Gamma=0.29$ eV, which is very similar to the $\Gamma=0.35$ eV case for the global fit.  These are both also similar to the IceCube-only best fit point, which is $\Gamma=0.40$ eV~\cite{Moulai:2021zey}.   Thus, this parameter appears to be consistent across data sets.

\subsection{3+2 and 3+3 fit Results}

One can question whether the improvement comes specifically from damping or from adding extra degrees of freedom in general.
Two often-tested models with substantially higher degrees of freedom are 3+2 and 3+3 models, with 7 and 12 parameters, respectively.
Unlike adding a damping term, these models add additional mass splittings that produce interference effects, allowing very complex waves to be fit to the global data.  

The tension results are reported on Table~\ref{table:fitquality}, where the asterisk (*) denotes that these fits do not include IceCube.
IceCube cannot be included in these fits because the necessary likelihood information is not released.
Therefore, we fit a separate set of 3+1 data without IceCube for comparison. 
The overall $\Delta \chi^2$/DOF measured with respect to the null is 56.9 (67.4)/ 7 (12) for 3+2 (3+3). 
For clarity, the parameters from the 3+2 fit are: the 2 mass splittings, 4 mixing elements ($U_{\alpha i}$), and 1 CP violating phase; and for the 3+3 fit are: the 3 mass splittings, 6 mixing elements ($U_{\alpha i}$), and 3 CP violating phases.

It is striking to note that the 3+2 tension is $2.0\sigma$ larger in these fits than for the 3+1 that includes IceCube, when one would expect that adding additional parameters would relieve tension.
This shows that the IceCube data is playing an important role and we encourage IceCube to pursue at least the 3+2 fits, recognizing that 3+3 fits may be computationally prohibitive. 

\section{Conclusion \label{sec:discuss}}
We have performed global fits to the 2022 short baseline vacuum-oscillation data sets and the $\nu_\mu$ matter resonance data from IceCube. The data show a 
preference for a 3+1+WP model or for a 3+1+dk model over only-3 or 3+1 models.
These models have the effect of damping oscillations at
low energy in the reactor data. Some modulation is preferred at high energies. These models, which each introduce one new parameter, are essentially similar in their major improvement to the fits. On the other hand, the 3+2 and 3+3 models increase and only minimally decrease tension in the fit, respectively.

These studies show that a large fraction of the tension between appearance and disappearance in 3+1 fits comes from the reactors, a point that has not previously been noted.
However, when the damped models are employed some tension remains.
The remaining tension is greatly improved when MiniBooNE is not included in the fit.
This may indicate that some fraction of the MiniBooNE excess events are inconsistent with the global picture of oscillations, as we have shown in this paper.

Lastly, we point out that the reactor event spectrum has a great deal more structure beyond the 5~MeV excess that is typically discussed.
Rector experiments have taken steps in analyzing data to reduce sensitivity to the unexplained structure. 
However, it is not clear that the techniques employed fully remove residual, $L$-dependent effects.
As a result, we urge the reactor community to revisit the systematic errors.
Results that set limits receive less scrutiny than results that indicate signals, and this might have led to systematic effects being overlooked.
However, it is crucial that this question be revisited because, if systematic effects have not been overlooked, then the structure in reactor experiments is pointing toward new physics that damps 3+1 oscillations.

\section{Acknowledgements}

MHS is supported by NSF grant PHY-1707971 and
NSF grant PHY-1912764 supported JMC, AD, and JMH.
NWK is also supported by the NSF Graduate Research Fellowship under Grant No. 1745302. 
CAA, IMS, and MJ are supported by the Faculty of Arts and Sciences of Harvard University.
Additionally, CAA and IMS are supported by the Alfred P. Sloan Foundation.  We thank B. Littlejohn for input on the PROSPECT results and I Shimizu, L.A. Winslow and A. de Gouv\^ea for discussions concerning KamLAND impact.

\bibliographystyle{apsrev}
\bibliography{global2022}

\pagebreak
\clearpage

\appendix

\section{More information on Fit Parameter for 3+1+WP \label{AppWP}}

\begin{figure}[ht]
 \includegraphics[width=0.32\textwidth]{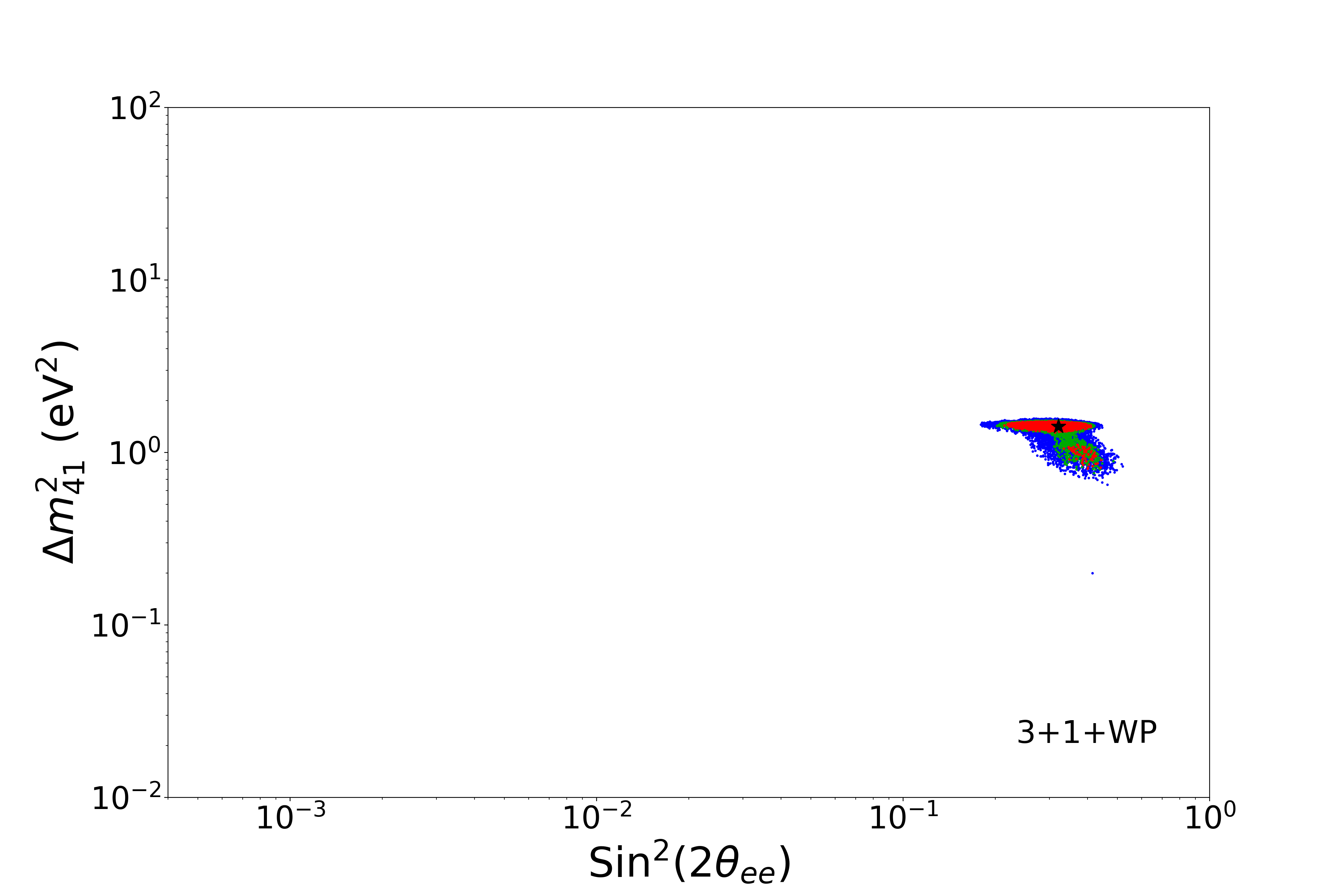}
  \includegraphics[width=0.32\textwidth]{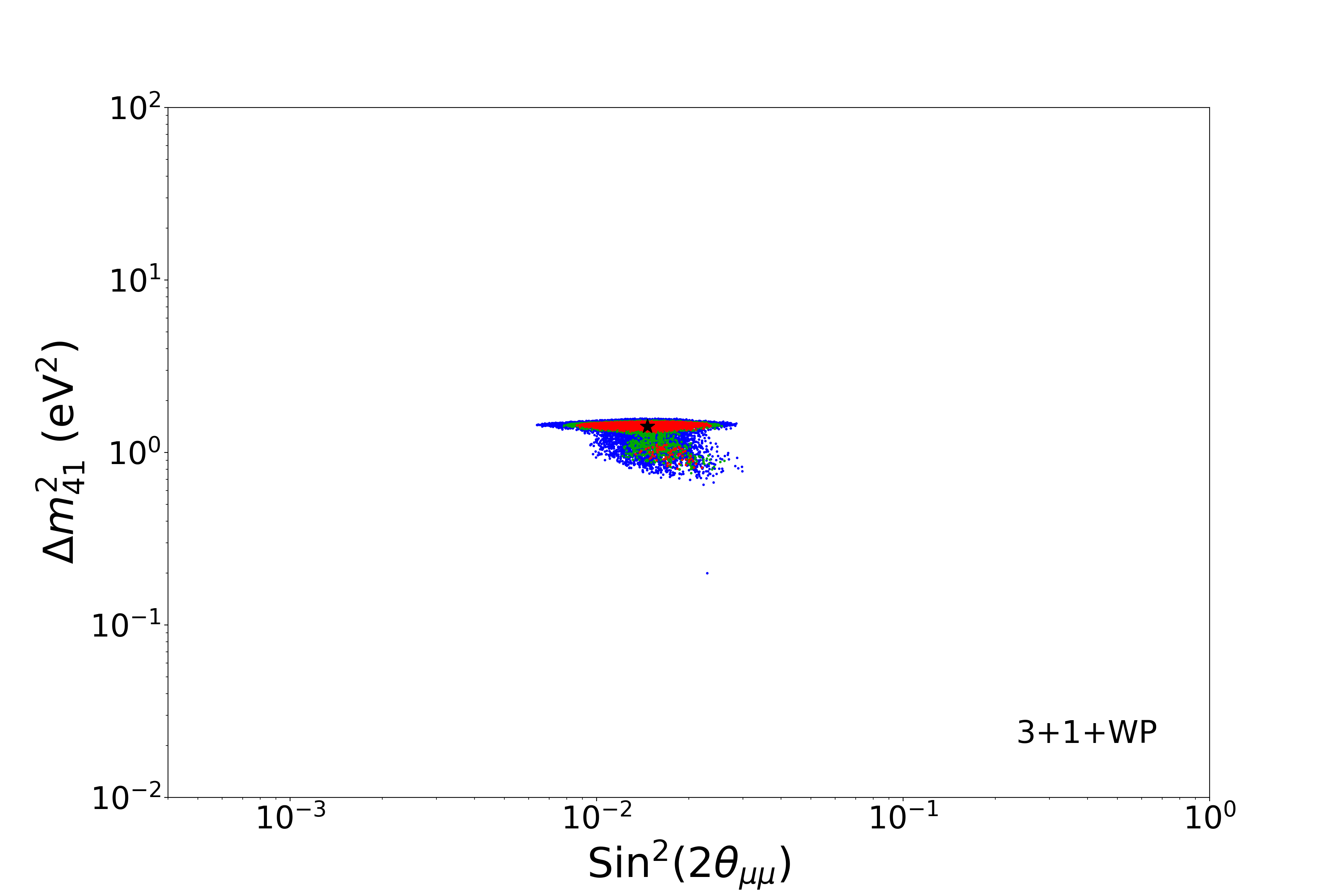}
   \includegraphics[width=0.32\textwidth]{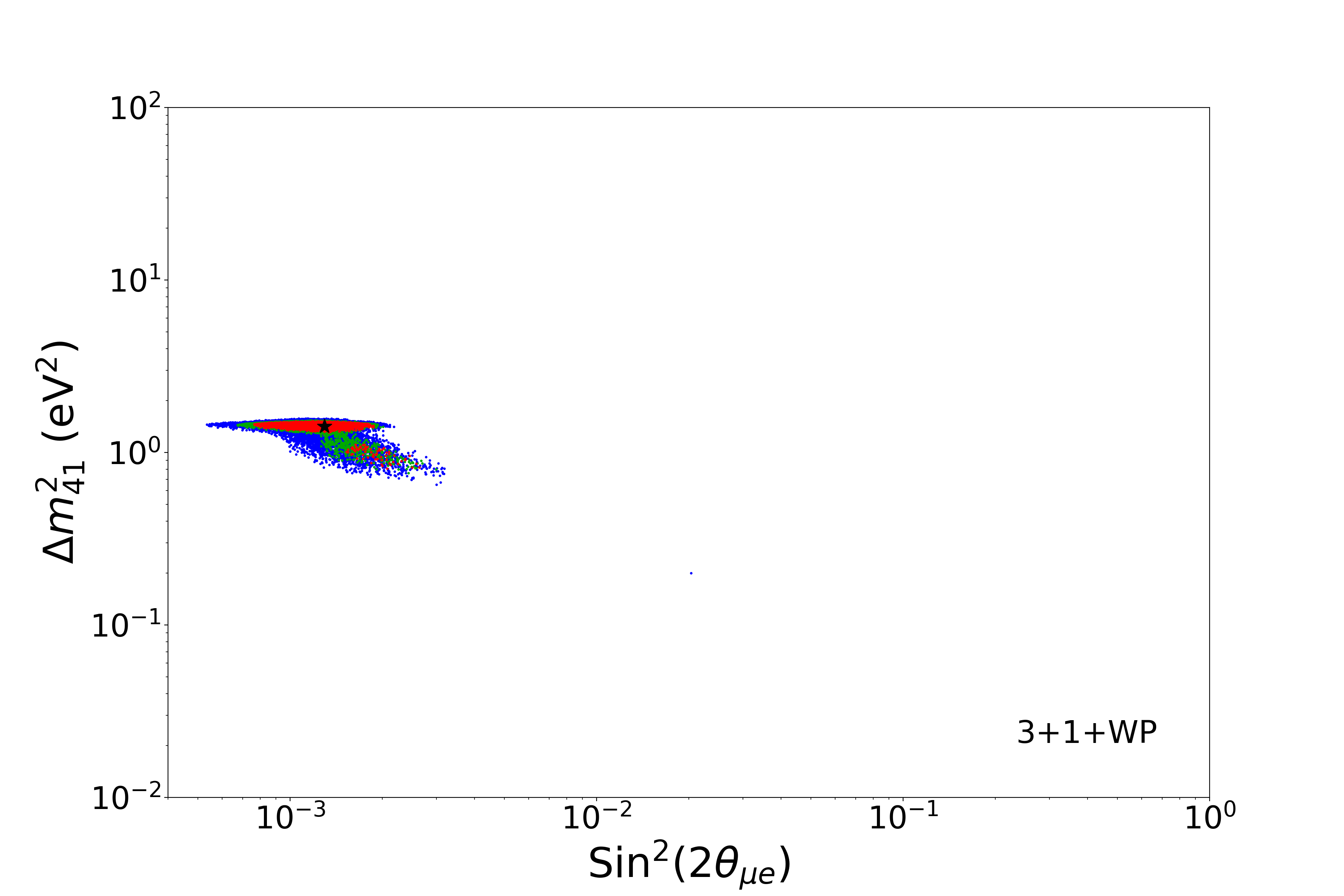}
\centering
\caption{\label{3+1+WPangles}  The 3+1+WP mixing angle and mass splitting space. Left:  $\nu_e$  disappearance; Center: $\nu_\mu$ disappearance; Right: appearance.  }
\end{figure}

For 3+1+WP, the projection profiled over the 4D parameters space for the three mixing angles is presented in Fig.~\ref{3+1+WPangles}.  The limits of the plots indicate the range of parameters explored in these fits.

\section{More information on Fit Parameters for 3+1+dk \label{Appdk}}

\begin{figure}[ht]
 \includegraphics[width=0.32\textwidth]{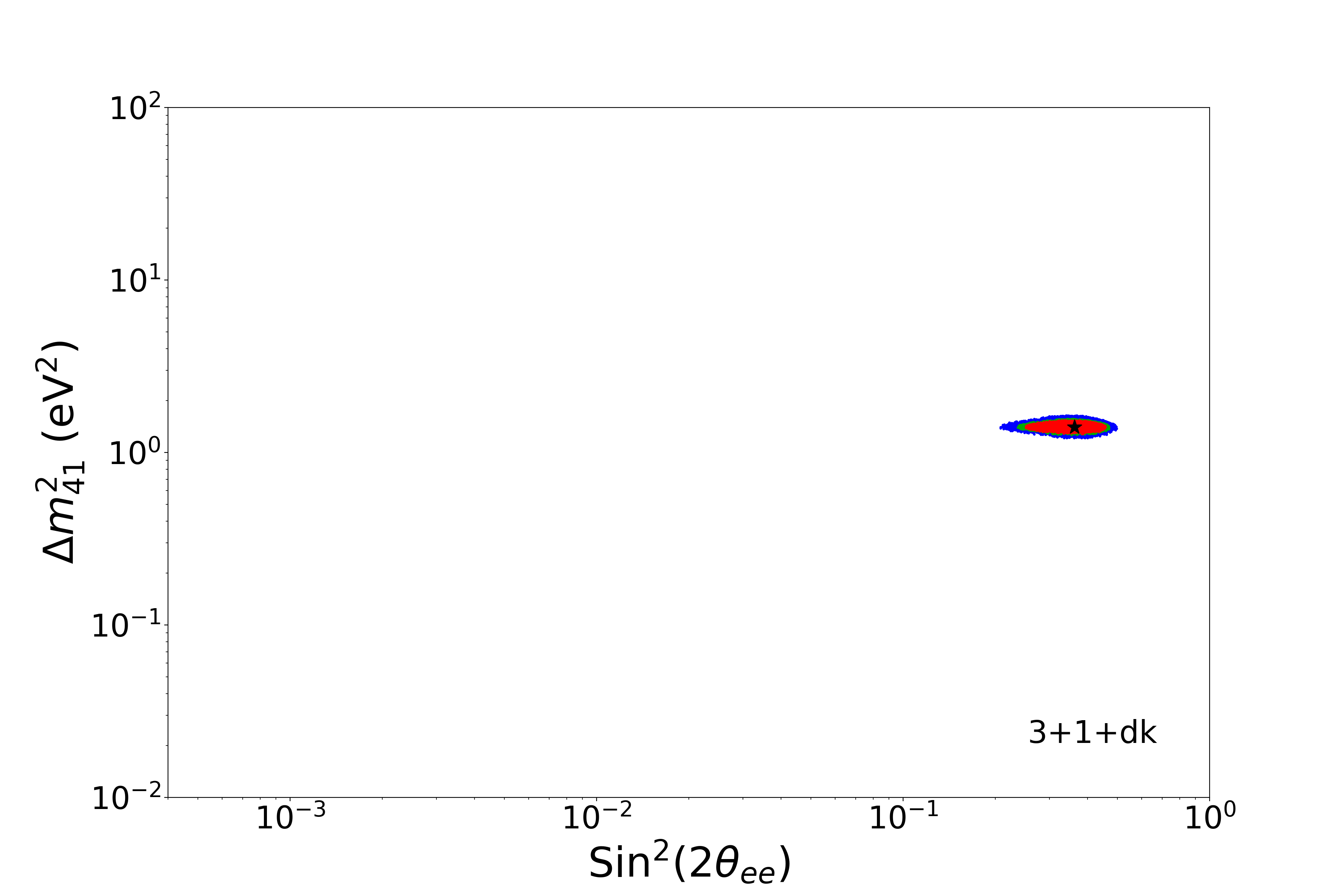}
  \includegraphics[width=0.32\textwidth]{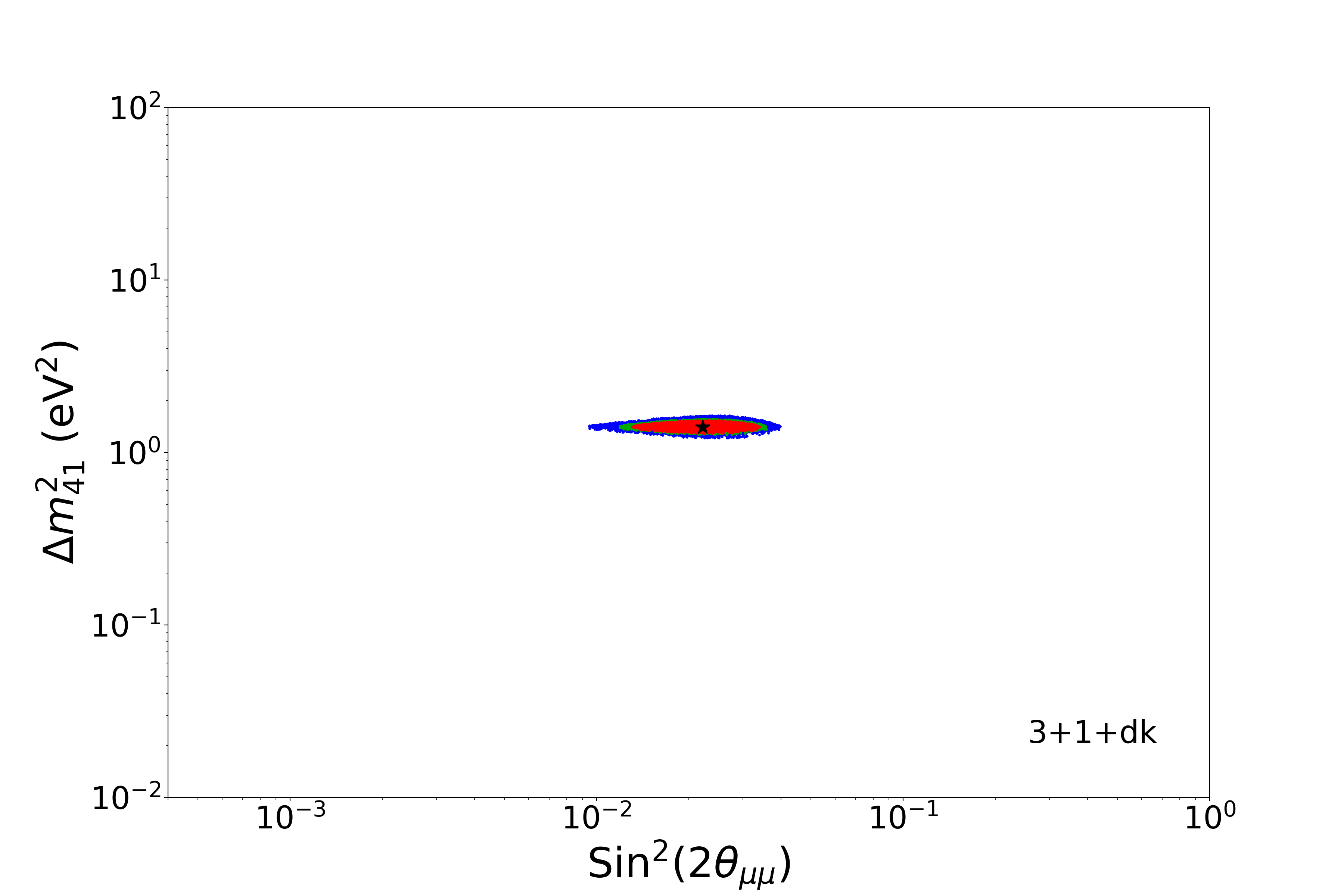}
   \includegraphics[width=0.32\textwidth]{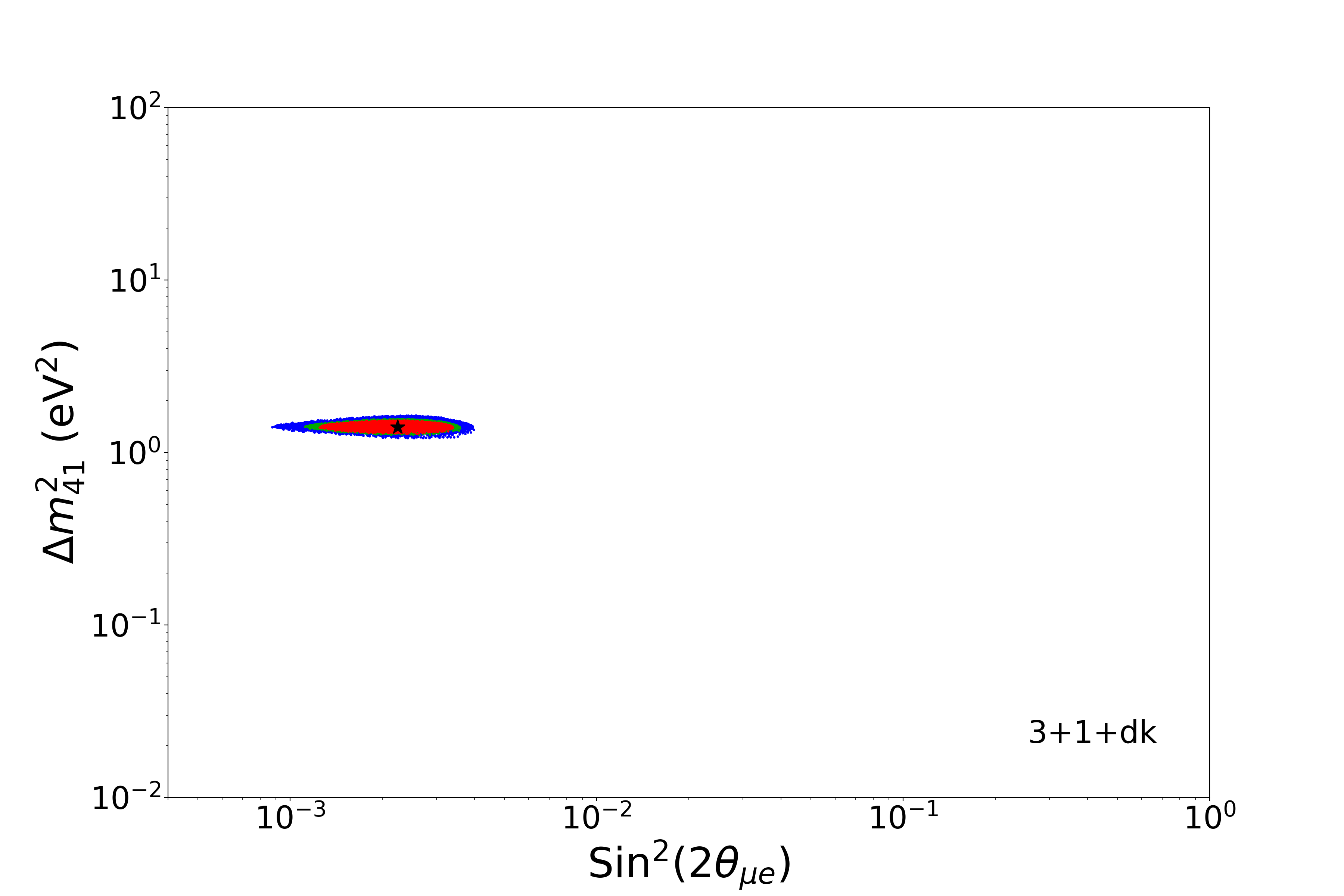} 
\centering
\caption{\label{3+1+dkangles}  The 3+1+dk mixing angle and mass splitting space. Left:  $\nu_e$  disappearance; Center: $\nu_\mu$ disappearance; Right: appearance.}
\end{figure}

\begin{figure}[ht]
     \includegraphics[width=0.5\textwidth]{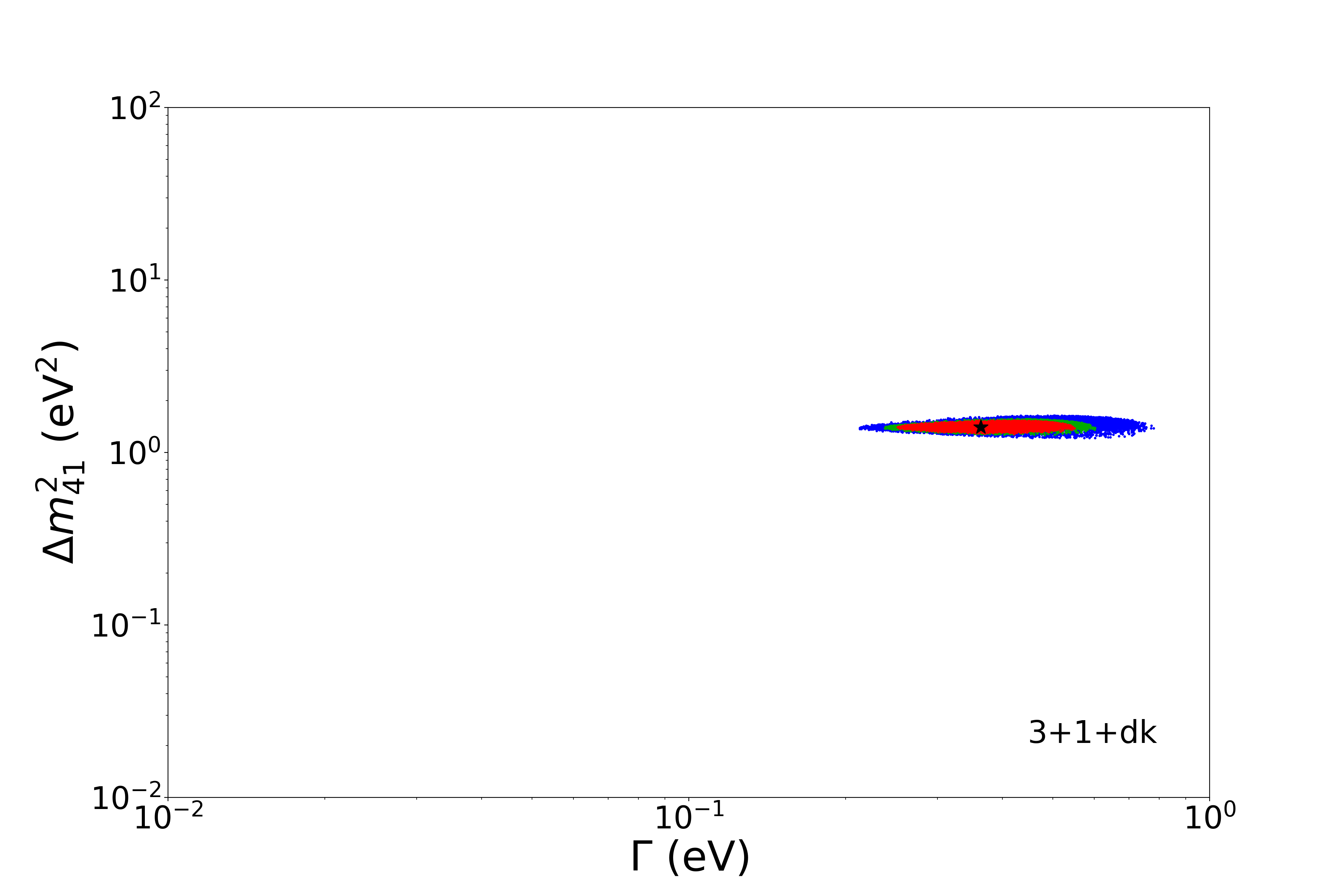}
\centering
\caption{\label{3+1+dkGamma}  The 3+1+dk $\Gamma$ and mass splitting space.}
\end{figure}

For 3+1+dk, the projection profiled over the 4D parameters space for the mixing angles and the mass splitting are presented in Fig.~\ref{3+1+dkangles}.   The decay width parameter space is explored in Fig.~\ref{3+1+dkGamma}.

\section{Fit Parameters for 3+2 and 3+3 \label{App3+N}}

Table~\ref{table:3+N} lists the best fit parameters for the 3+2 (top) and 3+3 (bottom) fits.

\begin{table}[ht]
\begin{center}
\begin{tabular}{| l | l |}
\hline
\multicolumn{2}{|c|}{3+2} \\ \hline
$\Delta m^2_{41}$, $\Delta m^2_{54}$ in eV$^2$ & 2.8E+1, 3.2E+1 \\
$U_{e 4}$, $U_{\mu 4}$ & 2.7E-1, 6.7E-1\\
$U_{e 5}$, $U_{\mu 5}$ &  1.6E-1, 1.2E-1\\ 
$\phi_{54}$ & 3.7E+0 \\ \hline
\multicolumn{2}{|c|}{3+3} \\ \hline
$\Delta m^2_{41}$, $\Delta m^2_{54}$, $\Delta m^2_{65}$ in eV$^2$ & 1.4E-2, 2.2E-1, 1.3E+1\\
$U_{e 4}$, $U_{\mu 4}$ & 7.3E-2, 9.8E-2\\
$U_{e 5}$, $U_{\mu 5}$ & 8.8E-2, 9.9E-1\\
$U_{e 6}$, $U_{\mu 6}$ & 2.8E-1, 2.7E-2\\ 
$\phi_{54}$, $\phi_{65}$, $\phi_{64}$ & 3.9E+0, 5.3E-1, 6.0E+0\\ \hline

\end{tabular}
\caption{\label{table:3+N}  Best fit parameters for 3+2 (top) and 3+3 (bottom). }
\end{center}
\end{table}

\end{document}

%% file: units.tex
\DeclareSIUnit \s {\second}
\DeclareSIUnit \ns {\nano\second}
\DeclareSIUnit \mus {\micro\second}
\DeclareSIUnit \ms {\milli\second}
\DeclareSIUnit \MB {\mega\byte}
\DeclareSIUnit \GB {\giga\byte}
\DeclareSIUnit \TB {\tera\byte}
\DeclareSIUnit \PB {\peta\byte}
\DeclareSIUnit \Mbps {\mega\bit/\s}
\DeclareSIUnit \Gbps {\giga\bit/\s}
\DeclareSIUnit \Tbps {\tera\bit/\s}
\DeclareSIUnit \Pbps {\peta\bit/\s}
\DeclareSIUnit \kton {\kilo\tonne} 
\DeclareSIUnit \kt {\kilo\tonne}
\DeclareSIUnit \kty {\kilo\tonne-\year}
\DeclareSIUnit \Mt {\mega\tonne}
\DeclareSIUnit \eV {\electronvolt}
\DeclareSIUnit \keV {\kilo\electronvolt}
\DeclareSIUnit \MeV {\mega\electronvolt}
\DeclareSIUnit \GeV {\giga\electronvolt}
\DeclareSIUnit \TeV {\tera\electronvolt}
\DeclareSIUnit \PeV {\peta\electronvolt}
\DeclareSIUnit \EeV {\exa\electronvolt}
\DeclareSIUnit \m {\meter}
\DeclareSIUnit \cm {\centi\meter}
\DeclareSIUnit \nm {\nano\meter}
\DeclareSIUnit \in {\inchcommand}
\DeclareSIUnit \km {\kilo\meter}
\DeclareSIUnit \kV {\kilo\volt}
\DeclareSIUnit \kW {\kilo\watt}
\DeclareSIUnit \MW {\mega\watt}
\DeclareSIUnit \MHz {\mega\hertz}
\DeclareSIUnit \mrad {\milli\radian}
\DeclareSIUnit \year {years}
\DeclareSIUnit \POT {POT}
\DeclareSIUnit \sig {$\sigma$}
\DeclareSIUnit\parsec{pc}
\DeclareSIUnit\lightyear{ly}
\DeclareSIUnit\foot{ft}
\DeclareSIUnit\ft{ft}
\DeclareSIUnit \ppb{ppb}
\DeclareSIUnit \ppt{ppt}
\DeclareSIUnit \samples{S}
\DeclareSIUnit \pe{PE}
\DeclareSIUnit \GeVmwe{GeV/mwe}
\DeclareSIUnit \mwe{mwe}

\newcommand{\enu}{\E_\enu}

%% file: tikz_style.tex
\usetikzlibrary{trees}
\usetikzlibrary{decorations.pathmorphing}
\usetikzlibrary{decorations.markings}

\definecolor{kyelloworange}   {RGB}{255, 210,  110}
\definecolor{jblue}  {RGB}{20,50,100}
\definecolor{npurple}  {RGB} {153, 51, 204}
\definecolor{wred}   {RGB}{217,0,56}
\definecolor{white}   {RGB}{255,255,255}
  
\definecolor{korange}   {RGB}{235, 80,  43}
\definecolor{korange2}   {RGB}{245, 100,  63}
\definecolor{kyelloworange}   {RGB}{255, 210,  110}
\definecolor{kyelloworange2}   {RGB}{240, 170,  90}
\definecolor{kred}   {RGB}{204,  102, 153}
\definecolor{kpurple}   {RGB}{153,  61, 190}
\definecolor{kpurplelight}   {RGB}{213,  161, 230}

\tikzset{
	    gaugeboson/.style={decorate,decoration={snake, amplitude = 3pt, post length = 1 pt, pre length = 4 pt},draw=magenta},
	    fermion/.style={draw=black,postaction={decorate},decoration={markings,mark=at position .55}},
	    fermionin/.style={draw=black,postaction={decorate},decoration={markings,mark=at position .55 with {\arrow[draw=black]{<}}}},
	    fermionout/.style={draw=black,postaction={decorate},decoration={markings,mark=at position .55 with {\arrow[draw=black]{>}}}},
	    gluon/.style={decorate,draw=magenta,decoration={coil,amplitude = 6pt,segment length=8pt}},
	    connect/.style={draw=black,postaction={decorate},decoration={markings}}, 
	    gluon2/.style={decorate,draw=magenta,decoration={coil,amplitude = 3pt,segment length=4pt}},
	    gaugeboson2/.style={decorate,decoration={snake, amplitude = 3pt, segment length=6pt},draw=black}
	}
	
\tikzset{
	  photon/.style={decorate, decoration={snake}, draw=npurple,very thick},
	  boson/.style={decorate, decoration={snake}, draw=npurple,very thick},
	  electron/.style={draw=jblue,very thick, postaction={decorate},
	           decoration={markings,mark=at position .55 with {\arrow[draw=jblue]{>}}}
	  },
	  electron2/.style={draw=jblue,very thick, postaction={decorate},
	           decoration={markings,mark=at position .55 with {\arrow[draw=jblue]{<}}}
	  },
	  fermion/.style={draw=jblue,very thick, postaction={decorate},
	            decoration={markings,mark=at position .55 with {\arrow[draw=jblue]{}}}
	  },
	  gluon/.style={decorate, draw=korange,very thick, 
	    decoration={coil,amplitude=4pt, segment length=6pt}},
	  higgs/.style={draw=wred,very thick, postaction={decorate},
	           decoration={markings,mark=at position .55 with {\arrow[draw=wred]{>}}}
	  },
	  nothing/.style={draw=white,very thick}
}